\documentclass[times,thesis]{uom_eee_dissertation_casson}

\usepackage{graphicx,psfrag,color} 
  \graphicspath{ {./images/} }
\usepackage{amsmath}               
  \allowdisplaybreaks[1]           
\usepackage{amssymb}               
\usepackage{url}                   
\usepackage{rotating}              
\usepackage{multirow}              
\usepackage{lscape}                
\usepackage{tabularx}              
\usepackage{verbatim}              
\usepackage{footnote}              
\usepackage{float}                 
\usepackage{booktabs}              
\usepackage{lipsum}                
\usepackage[position=top]{subfig}

\usepackage{algpseudocode}
\algrenewcommand\alglinenumber[1]{\scriptsize #1:}
\algrenewcommand\algorithmicindent{1.0em}
\usepackage{algorithm}
\usepackage{mathtools}
\usepackage{setspace}
\usepackage[sfdefault]{carlito}

\newcommand{\sus}[1]{$^{\mbox{\scriptsize #1}}$} 

\newcommand{\equ}[1]{(\ref{#1})}

\newcommand{\ieee}[1]{\citeauthor*{#1}~\cite{#1}}

\usepackage[style=ieee,backend=biber,backref=true,hyperref=auto]{biblatex}

\DefineBibliographyStrings{english}{backrefpage = {cited on p\adddot},  backrefpages = {cited on pp\adddot}}

\begin{document}

\maketitle

\begin{abstract} 
\noindent Smart grids are being increasingly deployed worldwide, as they constitute the electrical grid of the future, providing bidirectional communication between households. One of their main potential applications is the peer-to-peer (P2P) energy trading market, which promises users better electricity prices and higher incentives to produce renewable energy, while also increasing energy efficiency by reducing the electricity waste that arises from transportation across long distances. However, most P2P markets require users to submit energy bids/offers well in advance of their corresponding time slot, which cannot account for unexpected surpluses of energy consumption/production. Moreover, the fine-grained metering information used in calculating and settling bills/rewards is inherently sensitive and must be protected in conformity with existing privacy regulations. \\ \ \\
To address these issues, this report proposes a novel privacy-preserving billing and settlements protocol, PPBSP, for use in local energy markets with imperfect bid-offer fulfillment, which only uses homomorphically encrypted versions of the users’ half-hourly consumption data. In addition, PPBSP also supports various cost-sharing mechanisms among market participants, including two new and improved methods of proportionally redistributing the cost of maintaining the balance of the electrical grid in a fair and intuitive manner. \\ \ \\
An informal privacy analysis is performed in order to highlight the privacy-enhancing characteristics of the protocol, which include metering data secrecy and bill confidentiality. The performance-related properties of PPBSP are also evaluated in terms of both computation cost and communication overhead, demonstrating its efficiency and feasibility for markets ranging from the size of small local communities to large urban centres, thanks to the highly parallelizable nature of the billing algorithm and the low computational load placed on the users' smart meters.

\end{abstract}%
\clearpage
\uomtoc
\uomlot
\uomlof
\uomloa 
\uomabbr
\begin{table}[h]
\small
\hypertarget{table:abbrev}{}
\centering
\resizebox{\columnwidth}{!}{
\begin{tabular}{l  @{\hskip 0.3in}  p{0.6\linewidth}}
\textbf{ACD} & Aggregate consumption data \\
\textbf{BillCalc} & Bill calculation \\
\textbf{CD} & Consumption data \\
\textbf{ECD} & Encrypted consumption data \\
\textbf{FiT} & Feed-in tariff \\
\textbf{GridOp} & Grid operator \\
\textbf{HomoDec} & Homomorphic decryption \\
\textbf{HomoEnc} & Homomorphic encryption \\
\textbf{KeyGen} & Key generation \\
\textbf{LEM} & Local energy market \\
\textbf{MPC} & Multi-party computation \\
\textbf{P2P} & Peer-to-Peer\\
\textbf{P2PM} & Peer-to-Peer market \\
\textbf{PPBSP} & Privacy-preserving billing and settlements protocol \\
\textbf{RES} & Renewable energy source \\
\textbf{RM} & Retail market \\
\textbf{RP} & Retail price \\
\textbf{SG} & Smart grid \\
\textbf{SM} & Smart meter \\
\textbf{SOTA} & State of the art \\
\textbf{TD} & Total deviation \\
\textbf{TDD} & Total demand deviation \\
\textbf{TP} & Trading price \\
\textbf{TrPlat} & Trading platform \\
\textbf{TSD} & Total supply deviation \\

\end{tabular}
}
\end{table}
\clearpage
\uomnot
\begingroup
\renewcommand{\arraystretch}{1.1} 
\begin{table}[h]
\small
\hypertarget{table:not}{}
\centering
\resizebox{\columnwidth}{!}{%
\begin{tabular}{l  @{\hskip 0.15in}  p{0.6\linewidth}}
$\{X\}_{pub}$ & Value `X' encrypted using key `pub' \\
$pub\_X$, $priv\_X$ & Public key of `X', Private key of `X' \\
$U_n$ / $N_u$ & Total no. of users \\
$U^x$ & Individual user \\
$U_{val}$ & Individual meter reading of user \\
$U_{P2P}$ & Individual committed volume of user \\
$InDev_x$ & Individual deviation of a participant \\
$InDev_i$, $InDev_j$ & Individual deviation of a consumer / prosumer \\
$C_n$ & Total no. of Non-P2P consumers \\
$C_i$ & Individual consumer \\
$P2P^c_n$ & Total no. of P2P consumers \\
$P2P^c_{n,k}$ & Total no. of P2P consumers per supplier\\
$P2P^c_i$ & Individual P2P consumer \\
$C^{P2P}_{dem}$ & Individual committed demand of P2P consumer \\
$P_n$ & Total no. of Non-P2P prosumers \\
$P_j$ & Individual prosumer \\
$P2P^p_n$ & Total no. of P2P prosumers \\
$P2P^p_{n,k}$ & Total no. of P2P prosumers per supplier\\
$P2P^p_j$ & Individual P2P prosumer \\
$P^{P2P}_{sup}$ & Individual committed supply of P2P prosumer \\
$T^c_{over}$ & Total volume over-consumed \\
$T^c_{under}$ & Total volume under-consumed \\
$T^p_{over}$ & Total volume over-supplied \\
$T^p_{under}$ & Total volume under-supplied \\
$T\_up$ & Total volume under-consumed or over-supplied \\
$T\_down$ & Total volume over-consumed or under-supplied \\
$S_n$ / $N_s$ & Total suppliers \\
$N_{u,s}$ & Total no. of users per supplier \\
$S_k$ & Individual supplier \\
$S^{inc}_k$ & Individual supplier income \\
$S^{exp}_k$ & Individual supplier expenditure \\
$S^{bal}_k$ & Individual supplier balance \\
$S^{P2P}_k$ & Individual supplier P2P residue \\
$V^{RM}$ & Total volume traded at the retail market \\
\end{tabular}
}
\end{table}
\endgroup

\section{Introduction}
    \subsection{Motivation}
\begin{figure}[h]
          \centering
          \includegraphics[width=0.8\textwidth]{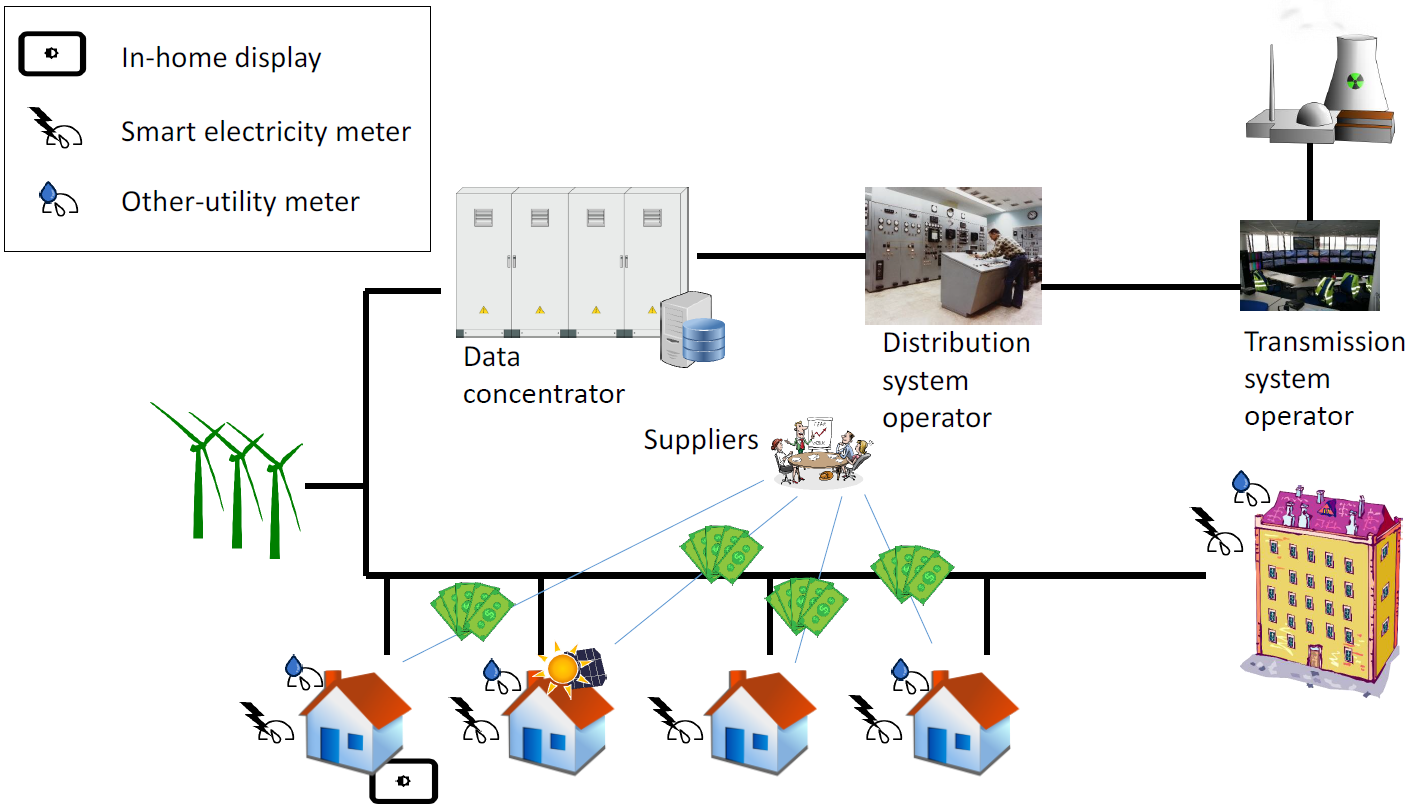}
          \caption{Illustration of a smart grid architecture~\cite{cleemput2018secure}.}
          \label{fig:smartgrid}
    \end{figure}

\begin{figure}[h]
          \centering
          \includegraphics[width=0.8\textwidth]{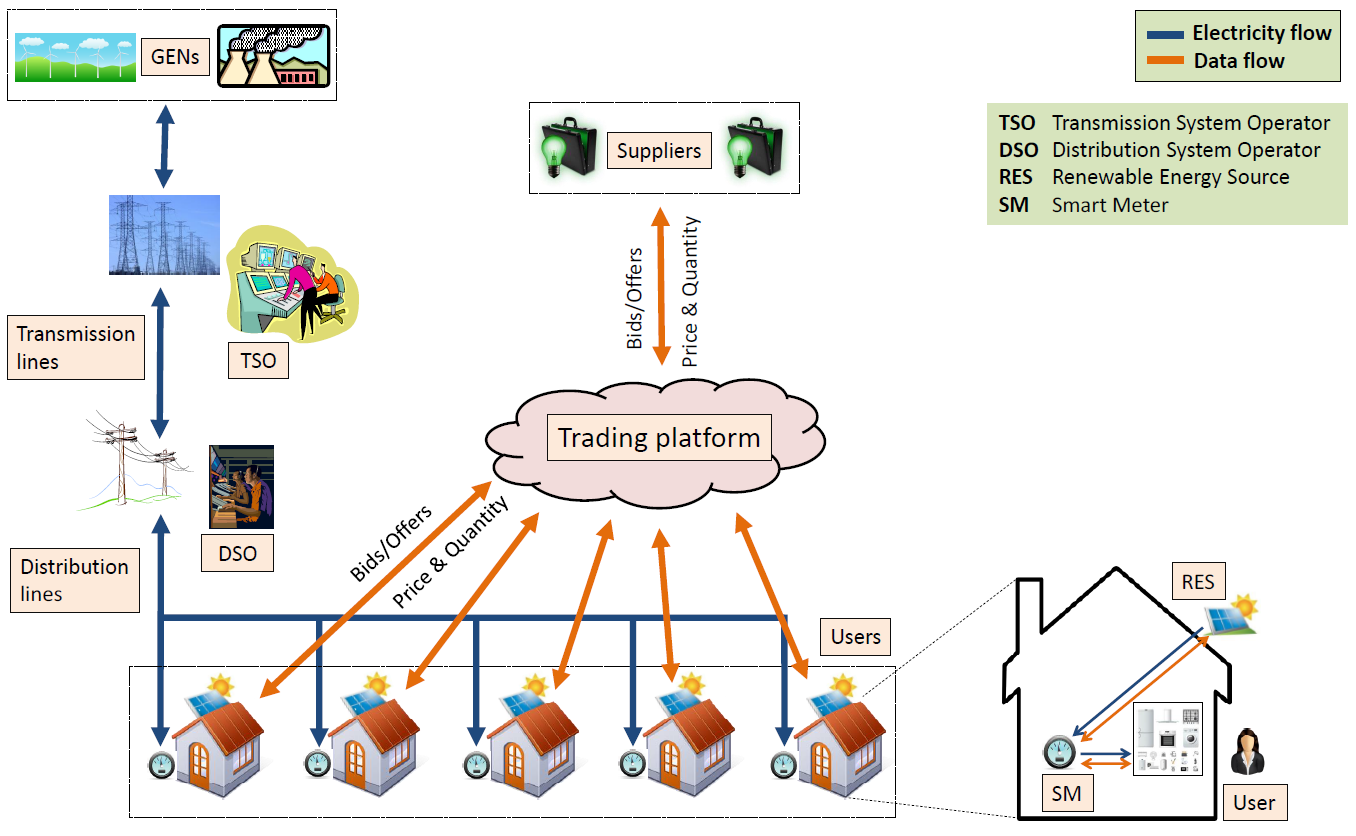}
          \caption{A proposed local energy trading market architecture~\cite{cleemput2018secure}.}
          \label{fig:lem}
    \end{figure}

    Many different researchers and stakeholders hold the vision that smart grid (SG) will be the next-generation electrical grid for providing electricity to millions of households around the world in a distributed manner~\cite{amin2005toward}. In essence, a smart grid is an advanced version of the traditional electrical grid system which also incorporates two-way electricity and communication flow capabilities between different devices that work together to collect, transmit, and analyze data in real time~\cite{fang2011smart}. A general smart grid architecture is illustrated in Figure~\ref{fig:smartgrid}. Among its various components, smart meters (SMs) are the ones to enable fine-grained, immediate monitoring of energy consumption and production,  supporting the integration of renewable energy sources and, most importantly, enabling the communication of such data for use by innovative SG applications~\cite{zheng2013smart}. With 29.5 million SMs already installed in UK households, representing 52\% of all electrical energy meters~\cite{kerai_2022}, it is planned that the automatic sending of such half-hourly meter readings from every single house and small business will become the default option by 2025~\cite{sm_2022}.\\
    \\The potential benefits of smart grids have attracted diverse stakeholders, including policymakers, energy providers, and also consumers. Governments and regulatory bodies recognise the importance of SGs for energy security, reducing greenhouse gas emissions, and increasing energy efficiency by reducing the waste of energy from transportation across long distances~\cite{farhangi2009path}. Meanwhile, energy suppliers and distributors view SGs as an opportunity to optimise their operations, reduce costs, and provide better customer service~\cite{bradley2013review}. A recent SG application that strongly benefits energy customers~\cite{karnouskos2011demand} and on which this report will focus is the peer-to-peer (P2P) electricity trading market~\cite{zhang2016bidding}, as the preferred implementation of the local energy market (LEM) model~\cite{teotia2016local}. The P2P market enables individuals or organizations to buy and sell electricity directly from each other, without the need for a centralized intermediary such as energy suppliers, leading to a more efficient and flexible energy market~\cite{schwidtal2022emerging}. Not only does it lead to better prices for consumers, but it also gives users a higher incentive to produce and sell their renewable energy, as the P2P trading price of electricity is highly advantageous to the fixed Feed-In Tariff (FiT) imposed by suppliers in traditional energy systems. Moreover, as the trading price at the P2P market is decided based on current supply and demand needs, the adoption of such a system has been shown to be capable of improving the local balance of energy generation and consumption~\cite{zhang2018peer}. An example LEM is illustrated in Figure~\ref{fig:lem}.
    \subsection{Problem Statement}
    In most P2P markets, users are required to submit bids/offers regarding their predicted energy consumption/production during a specific period (e.g. 30 minutes, one hour, etc.) ahead of time, using historical data of the household's load profile~\cite{hamlich2019short}, weather forecasts~\cite{fu2015using}, etc. Although accurate, it is impossible for any algorithm to predict the exact demand/supply of a household without fail, as unexpected surpluses of energy consumption/production are inevitable. This additional demand/supply represents the deviation between the user's commitments at the P2P market (their bid/offer) and their actual volume of energy consumed/produced during that period (measured by the SM). Together, these deviations can affect the efficient operation of the grid and increase the cost of keeping it balanced~\cite{dudjak2021impact}. \\
    \begin{figure}[h]
          \centering
          \includegraphics[width=0.8\textwidth]{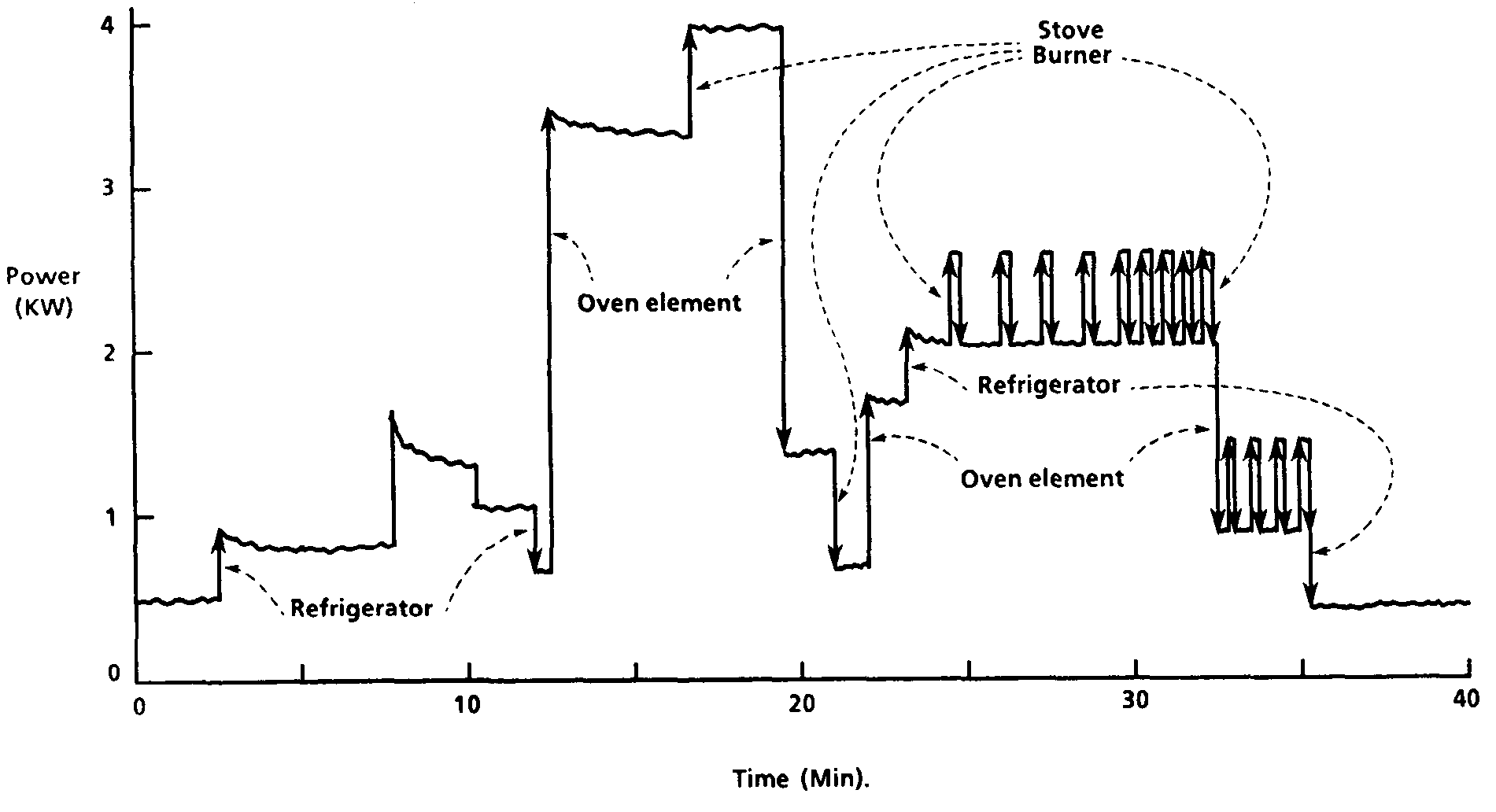}
          \caption{Non-intrusive appliance load monitoring example~\cite{192069}.}
          \label{fig:nilm1}
    \end{figure}
    \\Moreover, to enable the calculation and settlement of bills and rewards for users participating in the P2P electricity trading market, their fine-grained meter readings would need to be communicated to the trading platform, grid operators, suppliers, or even other participants in the market. Since this type of data inevitably contains sensitive information concerning the residents of a household~\cite{mustafa2016local, siddiqui2012smart}, and thus undoubtedly infringes both the European General Data Protection Regulation (GDPR) and the British Data Protection Act (DPA)~\cite{martinez2020smart}, particular attention has been paid to the possible private information that a malicious agent could discover about the inhabiting individuals. Among others, risks include the use of non-intrusive load monitoring (NILM) techniques~\cite{stankovic2016measuring} to infer the electrical consumption patterns of individual users, which could lead to the disclosure of privacy-invasive insights about specific medical conditions~\cite{fell2017energising}, religious beliefs, home appliance usage habits~\cite{greveler2012multimedia}, etc. Such, NILM is also one of the primary methods for performing consumer behaviour analysis~\cite{hart1989residential} and has the potential to lead to targeted advertising~\cite{gupta2017systems,leeb1996transient,haghighat2015system}, insurance adjustment, or discrimination and profiling~\cite{quinn2009privacy}. Figure~\ref{fig:nilm1} illustrates what type of information can be gathered from a household's electric power load profile. Therefore, the need arises for a privacy-upholding solution capable of working on a protected version of the user's sensitive data, including both the energy volume committed to the P2P market and the corresponding individual deviation, which would reveal only aggregated consumption data statistics to other curious market entities.

    \subsection{Aims and Objectives}
  To address these privacy concerns, the aim of this project is to design, implement and evaluate a novel privacy-preserving solution capable of calculating and resolving electrical energy bills and reward payments in local energy markets with imperfect bid-offer fulfillment, using only encrypted versions of the users' fine-grained consumption data. The solution should be also compatible with existing P2P trading markets from literature.\\ \ \\
  The objectives are outlined as follows: 
    \begin{itemize}
          \item Perform a review of existing literature in order to identify research gaps.
          \item Design a protocol capable of settling energy bills in a privacy-friendly manner for use in local energy markets.
          \item Implement and evaluate the proposed algorithm to verify its efficiency and practicality.
    \end{itemize}
    For protocol evaluation, an analysis of its privacy-related characteristics, as well as performance-related properties, will be conducted. First, the privacy and security analysis will examine whether the appropriate requirements established in the design preliminaries are upheld. Afterwards, the performance evaluation will test the computational complexity and the communication overhead of the proposed protocol, simulating the costs of real-world scenarios by varying the number of participating households, therefore covering a broad range of possible sizes for LEMs (from small, isolated communities to large metropolitan centres).

      \subsection{Limitations of Current Solutions}
    A privacy-enhancing solution for billing using a symmetric homomorphic encryption scheme has already been proposed by \ieee{alabdulatif2017}, while another privacy-friendly implementation of a local energy trading market using multi-party computation~\cite{abidin2016mpc}, later improved by the same authors through the addition of a simple billing algorithm has also been presented~\cite{abidin2018secure}. However, neither of these protocols takes into account the possible differences between the final meter readings and the volume that each user previously committed to trade for at the P2P market. \\ \ \\
    Despite the energy deviations' importance to the balance of the electrical grid, this topic has been mostly neglected in prior research, especially in the context of privacy-conscious implementations. The first to propose a solution for LEMs with imperfect bid-offer fulfillment, capable of bill adjustments in line with the respective individual deviations, were \ieee{thandi2022privacy}, whose protocol accomplishes the privacy-enhanced billing and settlements process using a partially homomorphic cryptosystem. However, the billing model presented only considers the individual deviations separately from each other, and thus inter-supplier aggregation of deviations is not supported, which would allow individual household deviations to compensate for each other, lowering energy costs for all P2P users. \\ \ \\
    While recent papers have proposed various billing models for P2P energy markets which take into account these half-hourly individual deviations, incentivising consumers/prosumers to minimise their own deviations by proposing different methods of distributing monetary rewards/punishments among the set of participants~\cite{madhusudan2022billing}, they were not designed to be at all privacy-preserving by themselves, but rather theoretical algorithms to be implemented in novel privacy-friendly protocols.
  
    \subsection{Contributions of the Work}
 To address the aforementioned limitations, we propose a novel privacy-preserving billing and settlements protocol (PPBSP) for local energy markets with imperfect bid-offer fulfillment. More specifically, the novel contributions of this work are fourfold:
 \begin{itemize}
 
    \item Design a novel privacy-protecting protocol, PPBSP, by improving on the current state-of-the-art (SOTA) solutions~\cite{thandi2022privacy}, which uses the Paillier partial homomorphic encryption scheme to calculate and settle bills in P2P markets, taking into account each user's individual deviation from the committed volumes and supporting the social splitting of their accompanying costs among all consumers/prosumers.
     \item Implement and improve two SOTA billing models from literature~\cite{madhusudan2022billing} by adding a proportional redistribution of the costs incurred for maintaining the grid's balance in the presence of individual deviations, leading to a more fair and intuitive calculation of bills and compensations for P2P market participants. 
     \item Perform privacy and security analysis of PPBSP in regard to internal system entities.
     \item Implement and evaluate the performance of PPBSP in terms of its computational and communication costs, both in theory (via analysis and extrapolation) and in practice (through simulations and measurements). The entire codebase is publicly hosted on my \href{https://github.com/Anndrey24/Privacy_Preserving_Algo}{personal GitHub account}.
     
 \end{itemize}
 The results of the work and research performed during project have been summarised and submitted in the form of a conference paper to the 14\sus{th} IEEE International Conference on Smart Grid Communications (SmartGridComm 2023).

    \subsection{Report Structure}
The rest of the report is organised as follows:
\begin{itemize}
          \item Section~\ref{sec:background} presents the necessary background knowledge and discusses related work.
          \item Section~\ref{sec:preliminaries} outlines the design preliminaries involved in implementing PPBSP.
          \item Section~\ref{sec:design} proposes and describes in detail the design properties of PPBSP.
          \item Section~\ref{sec:evaluation} analyses the privacy and security properties of the protocol and evaluates its performance in terms of computation and communication costs.
          \item Section~\ref{sec:conclusion} summarises my work and proposes future research avenues.
        \end{itemize}

\clearpage
\section{Background and Related Work} \label{sec:background}
    This section presents the necessary background information on smart grids and local energy markets, as well as the cryptographic building blocks used in the design of PPBSP. Afterwards, I also give an overview of the current research in privacy for smart grids, with an emphasis on settling bills in LEMs.
    \subsection{Local Energy Trading Concepts}
        \subsubsection{Smart Grid}
        The smart grid (SG) represents an enhancement of the traditional electrical grid infrastructure through the implementation of a two-way communication network connecting its various entities and components~\cite{Farhangi2010}. One of its main components is the smart meter (SM), a device which replaces traditional electricity meter on the customer's premises, capable of the following crucial functions:
        \begin{itemize}
            \item It is capable of measuring the flow of electricity in both directions, from the grid to the household and vice versa.
            \item It facilitates bidirectional communication with other actors within the SG system.
            \item It has the ability to measure various parameters related to the flow of electricity, including voltage level and frequency.
        \end{itemize}
        This integration of automatic meter readings can lead to a more reliable and efficient grid through automatic grid management systems and to more accurate and efficient billing and energy trading. The latter constitutes the motivation for the development of local energy market models.
        
        \subsubsection{Local Energy Market}
        The status quo in most liberalised energy markets limits customers to only buying or selling electricity from their energy supplier, thus leaving them few options to optimise their prices and providing no incentive to adopt renewable energy sources (e.g. solar panels), as any excess production is automatically injected back into the grid for little to no remuneration. \\
        \begin{figure}[h]
          \centering
          \includegraphics[width=0.8\textwidth]{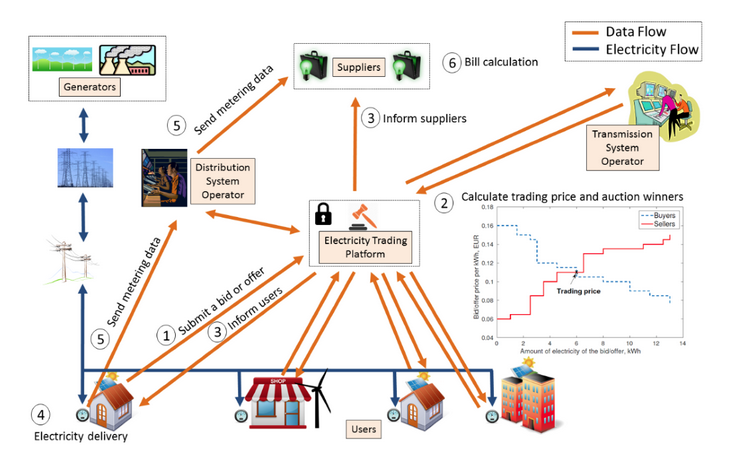}
          \caption{A P2P local electricity trading process~\cite{cleemput2018secure}.}
          \label{fig:lem_process}
        \end{figure}
        \\Contrary to this, LEMs permit users to trade electricity with other users and suppliers~\cite{mustafa2016local}. This means that a household generating more electricity than needed to cover its own demand is able to sell that surplus to another local consumer at the peer-to-peer trading market. Therefore, the P2P market participants set their own common trading prices, allowing them to purchase electricity at a lower price and also sell it at a higher one than the suppliers' offer. This process is illustrated in Figure~\ref{fig:lem_process}. The potential benefits of LEMs include the increased autonomy of microgrids, the decreased transmission-related electricity losses by encouraging the use of locally generated energy, and, most importantly, the additional incentives to install renewable energy sources. Naturally, for users who are unable to participate in the local electricity trading market, the supplier will remain an alternative source for both purchasing and selling electricity, thus serving as a secondary source for most households.
        
        \subsubsection{P2P Billing Model}
        Most P2P electricity trade markets determine the trading price through an auction mechanism in which the users participate by submitting their bids and offers in advance~\cite{CAPPER2022112403}. Therefore, an increased demand for electricity leads to a higher price, incentivising the prosumers to generate and sell more energy to compensate and the consumers to lower their consumption, whereas a larger supply than demand leads to a lower trading price, encouraging consumers to capitalise on and increase their consumption and prosumers to temporarily store away their electricity surplus. As this double auction mechanism takes place well in advance of the actual consumption/production time (ranging from 30 minutes to a day before), households are required to predict their future demand/supply values. As expected, such a process, however accurate, will lead to marginal, yet likely non-zero, differences between the volume that a user has committed to buy from/sell to the P2P market and their real meter readings, which are called individual deviations. \\ \ \\
        For a prosumer (a user that has sent an offer to sell electricity), a positive deviation means that the household supplied more energy than necessary, while a negative one denotes the under-production of electricity. Similarly, for a consumer (a user that has submitted a bid to purchase electricity), a positive deviation suggests consuming more than the committed volume, whereas a negative value indicates the unfulfillment of the household's consumption commitments. Moreover, in some instances, it is possible for a prosumer to under-supply to the extent that they actually become a net buyer of electricity for that trading period. The billing models proposed in this report are designed to seamlessly accommodate this case, too.\\ \ \\
        A billing model with imperfect bill-offer fulfillment takes into account both the user's committed volume to the P2P market and its individual deviation in order to incrementally calculate the monthly electricity bills by splitting the cost of the aforementioned deviations fairly among household~\cite{madhusudan2022billing}, using various methods that are further described in Section~\ref{sec:billing}. From beginning to end, the process of producing the bill for a single trading period (e.g. 30 minutes, one hour) consists of the following four steps:
        \begin{enumerate}
            \item Each user predicts the future demand/supply of the household using historical data, alongside other parameters, and submits the bid/offer to the P2P electricity auction.
            \item The trading price is decided by the relationship between current  supply and demand, while users are informed whether their bid/offer has been accepted by the auction.
            \item Meter readings are performed for the respective trading period and the discrepancies between them and the corresponding bid/offers are noted.
            \item Each user's partial bill is calculated by the trading platform.
        \end{enumerate}
        This work tackles the intricacies of the last two steps, from the moment real electricity consumption/production data is made available by the smart meter to the end of the bill settlement process.

    \subsection{Privacy Issues and Cryptographic Building Blocks}
        \subsubsection{Privacy Concerns of Smart Metering}
\begin{figure}[h]
          \centering
          \includegraphics[width=0.8\textwidth]{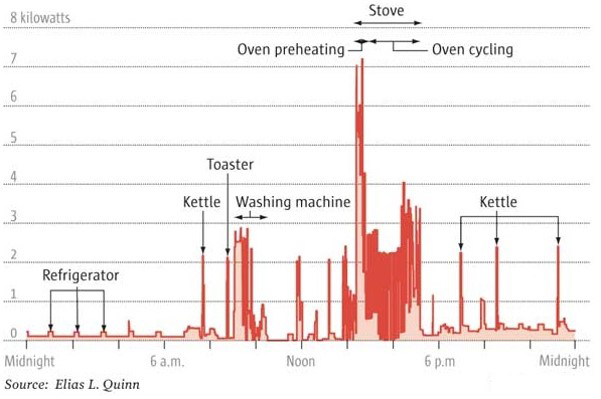}
          \caption{Recognising appliances in a detailed consumption pattern~\cite{newborough1999demand}.}
          \label{fig:nilm2}
        \end{figure}

        Multiple studies have shown that personal information can be inferred by observing a household's fine-grained electricity consumption data~\cite{4266955,lisovich2008,Bauer2009}. For example, non-intrusive load monitoring can reveal information regarding religion by observing early morning activity during the month of Ramadan, or provide health-related information like sleeping habits and cooking tendencies. Because of the unique nature of each appliance's load signature, it is also possible to identify specific appliances in a detailed consumption pattern~\cite{192069}, as illustrated in Figure~\ref{fig:nilm2}. This data can be used by various entities such as marketers, insurance companies, or criminals for reasons ranging from targeted advertisements and insurance adjustments to identifying the presence of valuable appliances and planning break-ins. 

        \subsubsection{Honest-but-Curious Model}
        This report extensively uses the honest-but-curious adversary model or semi-honest model~\cite{goldreich2004foundations}. This model assumes that parties follow the protocol accurately (i.e they are ``honest"), but they also actively seek to infer knowledge about other entities from all the inputs they receive or any intermediate computation results (i.e. they are ``curious"). Thus, honest-but-curious adversaries aim to maintain the proper functioning of the system in order to avoid being identified by monitoring mechanisms while also maximising their chance to infringe on others' privacy. \\ \ \\
        In the LEM case, honest-but-curious smart meters are trusted to only communicate accurate meter readings, the trading market will correctly calculate the users' bill, etc. However, suppliers, the market operator and other entities will also try to deduce individual load profiles by analysing values which have been submitted to them.
        
        \subsubsection{Cryptographic Schemes}
\begin{figure}[h]
          \centering
          \includegraphics[width=1\textwidth]{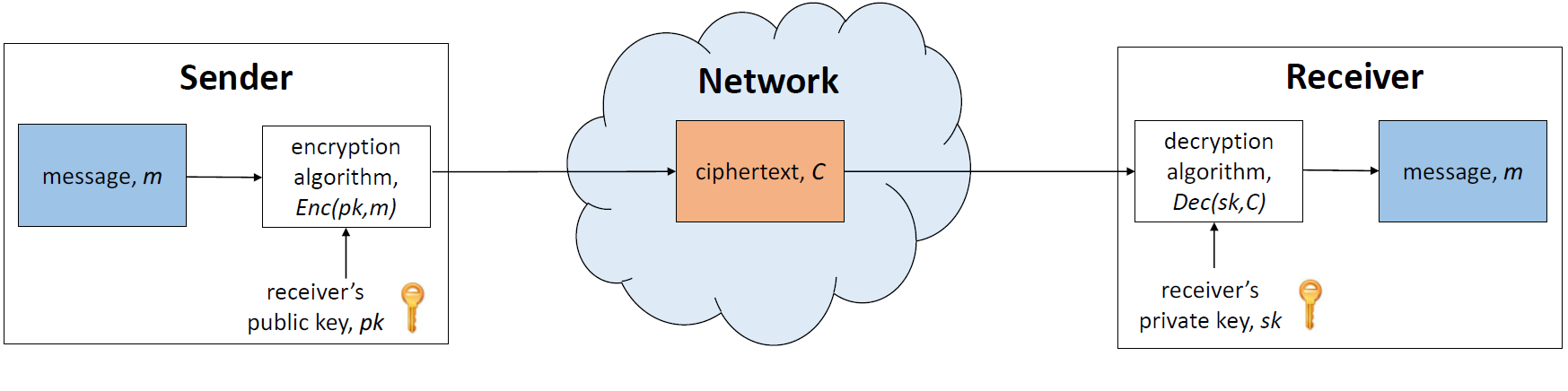}
          \caption{Asymmetric key cryptosystem~\cite{mustafa2015smart}.}
          \label{fig:assym}
        \end{figure}

        \paragraph{Asymmetric Key Encryption.}
        An asymmetric key cryptosystem involves the use of two distinct, but mathematically connected, keys by both the sender and the receiver in order to perform cryptographic operations, whose primary goals are message confidentiality, authenticity, etc. One of the keys, referred to as the public key, is made public, while the other key, known as the private key, must be kept confidential, accessible only by its owner~\cite{stallings_2005}. An example of such a cryptosystem being used to provide message confidentiality is depicted in Figure~\ref{fig:assym}.\\
        \paragraph{Homomorphic Encryption.}
        Homomorphic encryption represents a set of encryption functions which provide semantic security and enable specific algebraic computations to be carried out directly on the ciphertext, without the need for decryption. When the resulting ciphertext is finally decrypted, the output obtained is the same as that achieved if the operations were instead performed on the unencrypted data~\cite{rivest1978data}. These encryption schemes are mostly used in privacy-conscious solutions, in which mathematical operations must be carried out while keeping the inputs hidden. Homomorphic cryptosystems can be categorised into multiple types, the most important of which being:
        \begin{itemize}
            \item \textit{Partially homomorphic encryption (PHE)}: allows one select operation (addition or multiplication) to be performed on ciphertext an unlimited number of times.
            \item  \textit{Somewhat homomorphic encryption (SHE)}: allows two types of operations (addition and multiplication) to be performed on ciphertext a set number of times.
            \item  \textit{Fully homomorphic encryption (FHE)}: supports arbitrary computation on ciphertexts, being the most powerful type of homomorphic encryption, but limited by its large computational overhead.
        \end{itemize}
        
        \paragraph{Paillier Cryptosystem.}
        Out of the many homomorphic encryption schemes, the one that is of particular interest to this report is the Paillier cryptosystem~\cite{paillier1999public}. It is an efficient and semantically secure partially homomorphic cryptosystem, and, most importantly, it has an additive homomorphism property, desirable for privacy-preserving data aggregation. This means that multiplying the ciphertexts of any number of messages results in a ciphertext of the sum of all the messages, as described in \equ{equ:paillier}. Moreover, the cryptosystem is indeterministic, meaning that the same message will be encrypted in different resulting ciphertexts given distinct random values of the blinding factor $r$.\\ \ \\
        \begin{equation}
          \begin{aligned}
                C(m_1) \cdot c(m_2) 
                & = (g^{m_1} \cdot r^n_1) \cdot (g^{m_2} \cdot r^n_2) \mod{n^2} \\
                & = g^{(m_1+m_2)} \cdot (r_1\cdot r_2)^n \mod{n^2} \\
                & = C(m_1 + m_2)
            \end{aligned}
            \label{equ:paillier}
            \end{equation}
        \clearpage 
        The Paillier cryptosystem works as follows:
        \begin{itemize}
            \item Key Generation:
            \begin{enumerate}
                \item Select two large prime numbers $(p, q)$.
                \item Calculate $n = p \cdot q$ and $\lambda = lcm(p-1, q-1)$, where lcm means the least common multiple.
                \item Define the function $L(u) = (u-1)/n$.
                \item Choose a generator $g \in \mathbb{Z}^*_{n^2}$.
                \item Calculate $\mu = \left(L(g^\lambda \mod{n^2})\right)^{-1} \mod{n}$.
                \item $(n, g)$ is the public key.
                \item $(\lambda, \mu)$ is the private key.
            \end{enumerate}
            \item Encryption:
            \begin{enumerate}
                \item Given a message $m \in \mathbb{Z}_n$
                \item Select a random number $r \in \mathbb{Z}^*_{n}$.
                \item Compute ciphertext $C = Enc(m) = g^m \cdot r^n \mod{n^2}$.
            \end{enumerate}
            \item Decryption:
            \begin{enumerate}
                \item Given a ciphertext $c \in \mathbb{Z}^*_n$
                \item Decrypt it with $m = Dec(c) = L(c^\lambda \mod{n^2}) \cdot \mu \mod{n}$.
            \end{enumerate}
        \end{itemize}
        
    \subsection{Related Work}
    The preservation of users' privacy represents a topic of great importance to the feasibility of SG applications, which has been studied extensively by the research community~\cite{sultan2019privacy}. \ieee{mcdaniel2009security} are the first to identify the privacy-related vulnerabilities in SG systems and call for a broad national effort from government, academia and industry to propose and evaluate new solutions, while \ieee{kalogridis2013toward} present a unified framework that provides a methodological approach for integrating privacy into SGs. This report will focus only on a particular problem, which is the privacy of billing in local energy markets, instead of broadly covering the entire SG field. \\ \ \\
    A comprehensive security analysis of local energy trading markets was performed by \ieee{mustafa2016local}, raising several privacy threats and outlining a corresponding set of requirements for such a market. Various solutions partially addressing these concerns have been proposed. \ieee{uludag2015privacy} implemented a distributed bidding system that ensures the privacy of bidders, except for the winning bidder, whose identity is disclosed to the service provider, and later extended their design to accommodate a multi-winner auction mechanism~\cite{balli2017distributed}. \ieee{DENG2022108367} introduced an energy trading framework which does not expose the individual customer’s bidding price and volume by using homomorphic encryption. Multi-party computation (MPC)~\cite{yao1982protocols} has also been used to avoid privacy leakage. An energy auction mechanism implemented by \ieee{aly2017practically} allows generators and suppliers to trade electricity in an oblivious manner at the day-ahead market. However, none of these solutions describes in detail the  billing and settlements process which would happen after the trades are confirmed. \\ \ \\
    \ieee{8731} also present a detailed explanation of privacy threats that may arise in the context of SG implementations, as well as suggestions on how to address them. One of the proposed techniques for achieving privacy-preserving metering data aggregation for billing purposes involves the use of homomorphic encryption schemes, but it does not provide an in-depth analysis of this approach. Traditionally, homomorphic encryption has been used in the private aggregation of metering data for operational purposes~\cite{li2010secure, mustafa2015musp,mustafa2015dep2sa}, but it has also recently been used for private billing by~\cite{alabdulatif2017}. However, their implementation relies on a symmetric encryption scheme and thus requires a secure key exchange method to function. As for MPC, \ieee{abidin2016mpc} used it to design a privacy-friendly approach to local energy trading and later improved their protocol through the proposal of a simple privacy-preserving billing algorithm~\cite{abidin2018secure}. However, neither of these solutions takes into account the possible deviations between the final meter readings and the volume that each user committed to trade for at the P2P market. \\ \ \\
    Despite the significance of energy deviations in P2P markets, this issue has been largely overlooked in existing literature, with most prior research either assuming perfect fulfillment of the committed volumes~\cite{abidin2018secure}, neglecting to include these deviations into their billing models~\cite{wang2014game}, or introducing mechanisms which penalise market participants regardless of their personal contribution to the grid's energy imbalance~\cite{capper2022impact}. \ieee{thandi2022privacy} were the first to propose a homomorphic billing and settlements scheme which supports bill adjustments according to the individual differences in real electricity consumption/production compared to each user's initial trade commitments. However, the billing mechanism implemented only takes account of the individual deviations independently of each other and does not support any type of social splitting of the deviations' associated costs among P2P market participants, as proposed by \ieee{madhusudan2022billing}. \\ \ \\
    Unlike the aforementioned solutions, this report proposes a homomorphic privacy-preserving billing and settlements protocol for local energy markets with imperfect bid-offer fulfillment, which splits the incurred cost from the users' deviations fairly.

\clearpage
\section{Preliminaries} \label{sec:preliminaries}
This section outlines the system and threat model, assumptions and functional and privacy requirements used in the design of PPBSP.
\begin{figure}[h]
          \centering
          \includegraphics[width=1.0\textwidth]{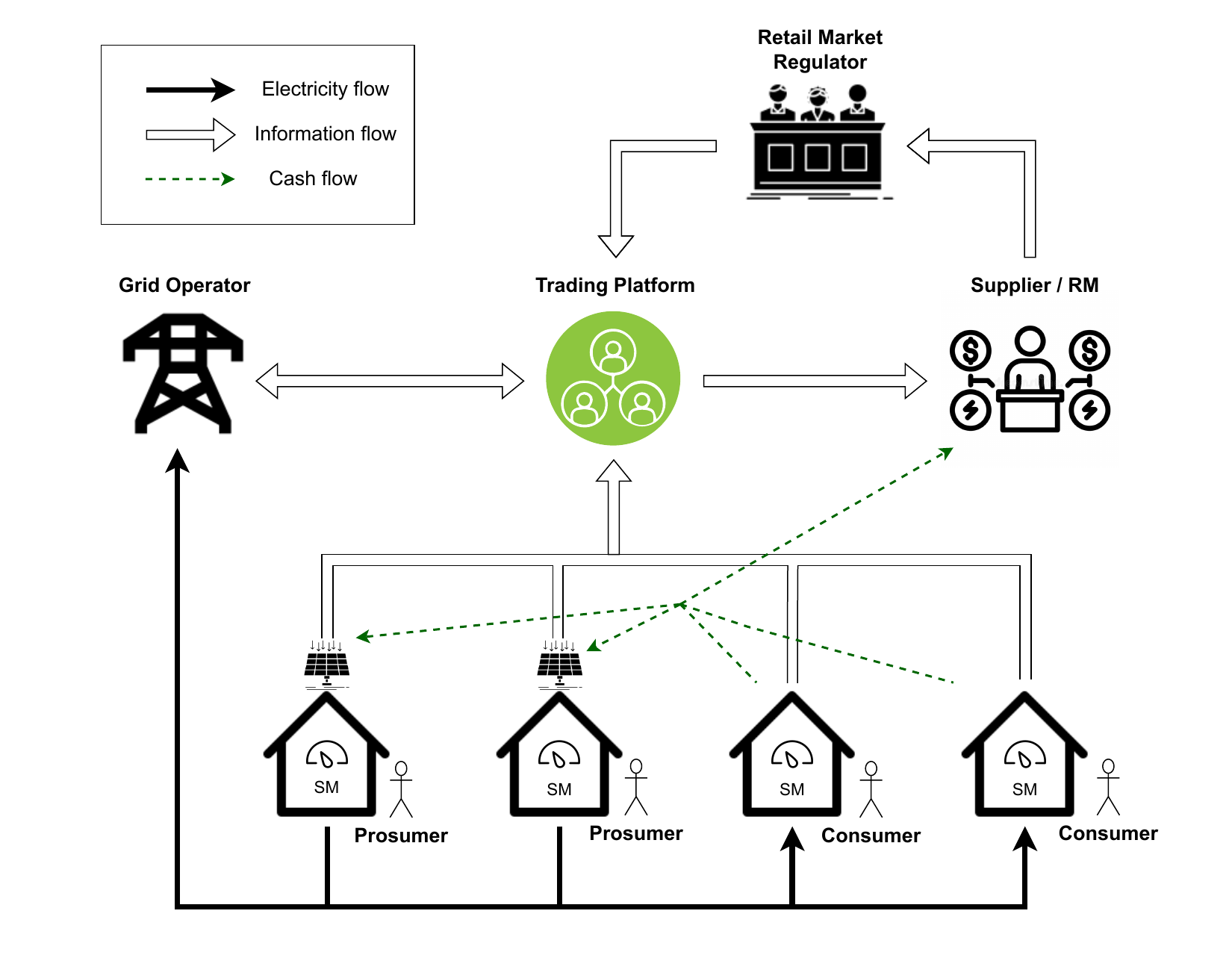}
          \caption{System model.}
          \label{fig:sysmodel}
\end{figure}
    \subsection{System Model}
    Figure~\ref{fig:sysmodel} illustrates the system model used in this report. Its consisting entities are the following: 
    \begin{enumerate}
	\item \textit{Users (Consumers / Prosumers)}:
            \begin{itemize}
            \item must pay monthly bills for the total energy consumption corresponding to their household;
            \item are compensated for their sold production of electricity from renewable energy sources (RES);
            \item can be either a net importer or net exporter of electricity in any given time slot;
            \item submit bids/offers for electrical energy using their smart meter (SM) based on predictions by their Home Energy Management Systems (HEMS);
            \item are temporarily classified as either consumers $C_i$ (buyers) or prosumers $P_j$ (sellers) for each time slot, based on their type of bid/offer submitted at the energy auction; and
            \item send actual half-hourly consumption/production data through their SM at the end of each trading period to calculate payments.
            \end{itemize}
        \item \textit{Peer-to-peer Market Operator / Trading Platform}:
        \begin{itemize}
            \item handles the energy double auction by accepting/rejecting bids and offers to clear the market; 
            \item sets the electricity trading price (TP) based on current supply and demand values; 
            \item keeps track of each user’s monthly net balance, updating after every trading period; 
            \item aggregates market-wide electricity usage statistics for each trading slot, to communicate to the grid operator;
            \item sends each supplier's net balance change for each trading period; and
            \item sends the final monthly bills of each user to its corresponding supplier.
        \end{itemize}
        \item \textit{Suppliers}:
        \begin{itemize}
            \item sell electrical energy to consumers at a retail buy price (RP);
	      \item buy electrical energy from prosumers at a fixed Feed-in Tariff (FiT);
            \item receive customer bills, including both the capital they traded at the P2P market (with other users) and the backup retail market (directly with the supplier); and
            \item keep the respective supplier's profit and send the rest to the other suppliers, distributed in accordance with the market regulator's guidance.
        \end{itemize}
        \item \textit{Grid Operator}:
        \begin{itemize}
            \item decrypts the aggregated energy usage statistics for each trading slot;
            \item communicates the decrypted values back to the P2P trading platform for their use in bill calculation; and
            \item serves as an independent entity capable of periodically (or on-request) checking the sincerity of the suppliers.
        \end{itemize}
        \item \textit{Retail Market Regulator}:
        \begin{itemize}
            \item coordinates the fair redistribution between suppliers of capital acquired by the supplier base in its role as an intermediary of the customer balances traded at the P2P market.
        \end{itemize}
    \end{enumerate}
	
    \subsection{Threat Model}
The threat model of the proposed solution is as follows:
 \begin{itemize}
     \item All entities presented in the system model are considered honest-but-curious entities, meaning that they will only follow the protocol and ruleset provided to them, but will also try to learn as much as possible from the information they have available~\cite{paverd2014modelling}. Therefore, the peer-to-peer market operator can be trusted to accurately compute the monthly bills and rewards, but cannot be allowed to have access to the raw smart meter data due to privacy concerns. Similarly, the grid operator is expected to provide correct decrypted values of the market deviation statistics, yet it must not be presented with non-aggregated user consumption data. \\ \ \\
     The energy suppliers represent a special case of semi-honest agents, as even if they were to behave as malicious entities capable of communicating erroneous, beneficial information about their customers' bills, the existence of an overseeing power makes getting caught and subsequently punished an inevitability, leaving the suppliers no rational motive for acting in a self-serving manner. Therefore, by circumstance, energy suppliers can also be modelled as honest-but-curious entities, trusted to authentically settle bills and payments as long as their truthfulness can be verified by another supervisor agent. 
     \item All external entities operating on the network are malicious. They will try to intercept, read and modify communications between the aforementioned agents.
 \end{itemize}

    \subsection{Assumptions}
    We use the following assumptions in our design: 
        \begin{itemize}
            \item The distributed system is supported by an underlying network that follows secure communication protocols, such as a robust public-key cryptosystem and digital signatures. Therefore, all communication is assumed to be safe from interception, reading, or modification by malicious entities. Thus, this report is only concerned with protecting sensitive information from its intended recipients.
            \item A user’s household, comprising of a smart meter (SM) and a Home Energy Management System (HEMS), is considered a single, temper-proof entity. The data it collects and sends is assumed to be correct.
            \item Every consumer/prosumer is assumed to pay their respective bill at the end of the billing period. In the event that they do not, the suppliers would shoulder the debt and deal with the situation accordingly.
            \item Each participant in the protocol has access to everyone’s homomorphic public keys and their own private key.
            \item There exists a double auction mechanism that sets a single unique trading price for both selling and buying electricity at the P2P market in line with the current supply and demand values.
            \item The TP is always valued between the FiT and the RP.
            \item The user is informed of the acceptance status of their bid/offer at the energy auction before the end of each trading period.
            \item The volume bought at the P2P market is equal to the volume sold at that same P2P market.
        \end{itemize}

	\subsection{Privacy and Security Requirements}
 Our designed solution should satisfy the following security and privacy requirements: 
         \begin{itemize}
             \item \textit{Metering data confidentiality}: The fine-grained (half-hourly) energy meter readings must only be known by the respective users. For all other uses, such as aggregation or bill calculation, the data must only be available in an encrypted format.
             \item \textit{Partial bill confidentiality}: The partial bill will be stored by the peer-to-peer market operator only in an encrypted format, which it is unable to decrypt by itself, until the end of the month in order to calculate the final monthly bill. That final bill can be shown to suppliers in order to settle bills as it no longer contains any privacy-sensitive information.
             \item \textit{Supplier accountability}: An authorised enquiring entity should be able to verify the veracity of each supplier's reported leftover capital after settling bills with its customers.
         \end{itemize}

\clearpage
\section{Protocol Design} \label{sec:design}
\begin{figure}[h]
          \centering
          \includegraphics[width=1.0\textwidth]{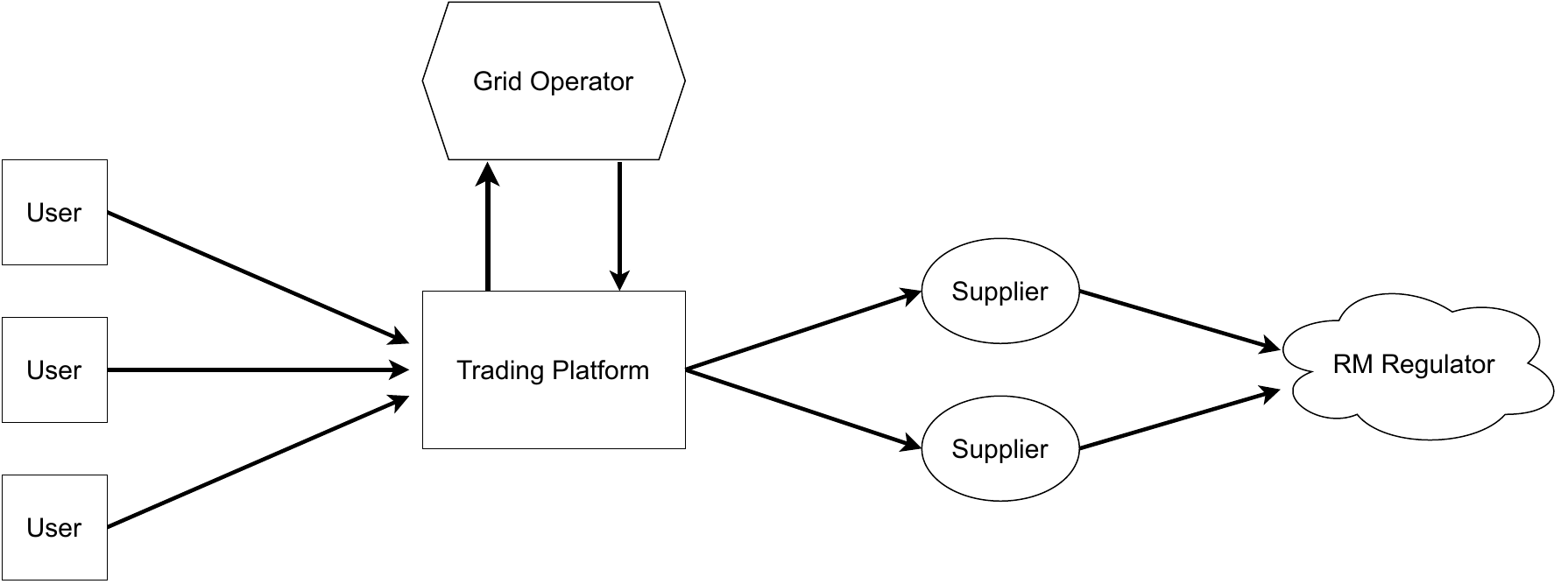}
          \caption{PPBSP overview.}
          \label{fig:PPBSP}
    \end{figure}

    This section details the novel privacy-preserving billing and settlement protocol PPBSP, which improves upon the SOTA solution for bill settlements in LEMs with imperfect bid-offer fulfillment~\cite{thandi2022privacy}. Before an intricate explanation of each entity's role and function, a general overview of the protocol is outlined first, explaining the separate phases of the algorithm, from system initialisation, to partial bill calculation, and finally bill settlement. Figure~\ref{fig:PPBSP} illustrates the general information flow of PPBSP.
    \subsection{Overview of PPBSP} \label{sec:design_overview}
        PPBSP comprises three distinct phases: system initialisation, partial bill calculation, and final bill settlement. A Python implementation is also publicly available on my \href{https://github.com/Anndrey24/Privacy_Preserving_Algo}{personal GitHub account}.  \\ \ \\
        Phase 1 is equivalent to the \textit{system initialisation}. In this phase, the energy suppliers, as well as the grid operator, generate a public-private homomorphic key pair, making the former available to any authorised enquiring entity, while keeping the latter only to themselves. Parallel to this, the P2P trading platform is initialised with a single corresponding grid operator, a list of users, identified using unique IDs, a list of energy suppliers, and a many-to-one relationship mapping each user to its respective supplier. \\ \ \\
        Phase 2 occurs at the end of each \textit{trading period}, e.g. every 30 minutes in the case of half-hourly meter readings, and represents the partial bill calculation, performed using the following steps. Each user's SM measures the current meter reading for the specific trading slot, separating the electricity consumption/production into committed energy amount and individual deviation, based on its prior bid/offer at the energy auction. Using this metering data along with the knowledge of its latest auction bid and access to a network of public keys, the SM constructs a final payload including homomorphically encrypted versions of the sensitive metering data and additional non-sensitive metadata, which is then sent to the trading platform. Using the received payloads, the trading platform calculates each user's partial bills for the corresponding trading period in accordance with one of the four implemented billing models, with the grid operator acting as a crucial part in market-wide data aggregation in two of the algorithms. After the partial bill calculation is complete, suppliers are sent aggregated data about their customers for that trading period, particularly the respective supplier's net balance change over that slot after transactions with its customers at the retail market. \\ \ \\
        Phase 3 is repeated after each \textit{billing period}, e.g. every month, and represents the final bill settlement, when the following steps take place. The trading platform aggregates each user's partial bills over the billing period and sends the final number to its corresponding energy supplier. Therefore, each supplier receives a list of final bills of their associated customers. At the same time, each supplier sums up their own net balance changes incurred in each trading period to arrive at a final net balance change value for that billing period. Acting as an intermediary of payments to the P2P market and the RM, after collecting payments or paying out rewards to its customers according to the final bills, each supplier subtracts its own earnings at the retail market (the final net balance) to keep to itself, with what remains being the residue of user transactions at the P2P market. These leftover differences (some positive, in the case where a supplier's customers bought more energy at the P2PM, and some negative, where a supplier's customers sold more energy at the P2PM) are communicated to the retail market regulator, which coordinates their fair redistribution such that the final aggregated residues from all suppliers equal 0, signifying a correct bill settlement process.\\
        \begin{figure}[h]
              \centering
              \includegraphics[width=0.8\textwidth]{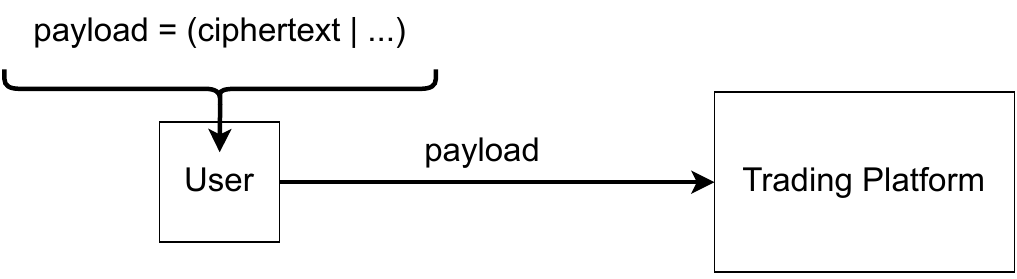}
              \caption{User protocol.}
              \label{fig:user}
        \end{figure}
    \subsection{User Protocol} \label{sec:user}
    At the end of every trading period, each user's SM is entrusted with the encryption and communication of the fine-grained metering data representing the electrical energy consumed/produced by that household over that slot (e.g. 30 minutes, one hour, etc.), in conformity with an accepted convention, in the form of a single payload containing all the necessary information for the trading platform to process and calculate the respective partial bills. This process is outlined in Figure~\ref{fig:user}. Moreover, PPBSP does not rely on the P2P trading platform having any pre-existing knowledge of the results of the energy auction. This design choice ensures a more distributed delegation of responsibilities across the system's entities and facilitates the protocol's integration with already existing double auction mechanisms from the literature~\cite{abidin2016mpc}, being an inherently modular approach. However, this means that all information regarding the acceptance status of user bids, their type (i.e. whether they were for buying or selling energy), as well as the energy amount which the respective user committed to trading at the P2P market must be communicated individually to the platform to allow for the calculation of the partial bills. \\ \ \\
    The step-by-step process of payload generation by the SM is presented in Algorithm~\ref{algo:user} and described below:
    \begin{algorithm}[t]
        \caption{User Protocol}\label{algo:user}
        \begin{algorithmic}[1]
            \setstretch{1.25}
            \Procedure{Payload}{$is\_bid\_accepted, bid\_type, U^{x}_{P2P}, U^x_{val}$}
                \State \hspace{\algorithmicindent}$net\_consumption\_time$ = $sign(U^x_{val})$
                \State \hspace{\algorithmicindent}$InDev_x$ = $\left(bid\_type \times U^x_{val}\right) - U^{x}_{P2P}$
                \State \hspace{\algorithmicindent}$\{U^x_{P2P}\}_{pub\_S_k}$ = $Enc(U^x_{P2P}, pub\_S_k)$
                \State \hspace{\algorithmicindent}$\{U^x_{P2P}\}_{pub\_GridOp}$ = $Enc(U^x_{P2P}, pub\_GridOp)$
                \State \hspace{\algorithmicindent}$\{InDev_x\}_{pub\_S_k}$ = $Enc(InDev_x, pub\_S_k)$
                \State \hspace{\algorithmicindent}$\{InDev_x\}_{pub\_GridOp}$ = $Enc(InDev_x, pub\_GridOp)$
                \State \Return $payload = (is\_bid\_accepted \| bid\_type \| net\_consumption\_type \| sign(InDev) \|$ \\
                \hskip3cm$ \{U^x_{P2P}\}_{pub\_S_k} \|\{InDev_x\}_{pub\_S_k} \| \{U^{x}_{P2P}\}_{pub\_GridOp} \|   \{InDev_x\}_{pub\_GridOp} )$
            \EndProcedure
        \end{algorithmic}
    \end{algorithm}
        \begin{enumerate}
          \item It reads the real energy consumption data for the specific trading period $U^x_{val}$, which then needs to be expressed in terms of: 
          \begin{itemize}
          \item whether or not the user's bid at the P2P market has been accepted $is\_bid\_accepted$:
          \begin{equation}
                is\_bid\_accepted =  
                 \begin{cases*}
                    1 & if bid accepted, \\
                    0 & otherwise
                    \end{cases*}
              \end{equation}
          \item whether the user offered to buy or sell energy during the auction (consumer/prosumer) $bid\_type$:
              \begin{equation}
                 bid\_type =  
                 \begin{cases*}
                    1 & if buying P2P energy, \\
                    -1 & otherwise
                    \end{cases*}
              \end{equation}
          \item whether the user ended up being a net buyer or a net seller of electricity over the trading slot $net\_consumption\_type$, equivalent to the sign of $U^x_{val}$:
                 \begin{equation}
                    net\_consumption\_type =  
                    \begin{cases*}
                        1 & if net buyer of energy, \\
                        -1 & otherwise
                    \end{cases*}
                \end{equation}
                \begin{center}
                    $\equiv$
                \end{center}
                \begin{equation}
                    net\_consumption\_type = sign(U^x_{val})
                \end{equation}
          \item the energy volume amount committed to the P2P market $U^{x}_{P2P}$:
            \begin{equation}
                U^{x}_{P2P} \in \mathbb{R}_{+}, \text{ according to the submitted bid}
            \end{equation}
          The committed value is always a positive number $U^{x}_{P2P} \geq 0$. Whether the volume was bought or sold at the P2P market is indicated by $bid\_type$.
          
          \item the individual deviation $InDev_x$, as a measure of the difference between the real meter reading and the volume committed to trading at the P2P market:
          \begin{equation}
                InDev_x = bid\_type \times U^x_{val} - U^{x}_{P2P}
            \end{equation}
            \begin{center}
                because
            \end{center}
            \begin{equation}
                 U^x_{val} = bid\_type \times \left(U^{x}_{P2P} + InDev_x\right), \text{where } bid\_type \in \{\pm1\}
            \end{equation}
          \item the sign of the individual deviation $sign(InDev_x) \in \{\pm1\}$. \\
          Paillier's cryptosystem does not natively support comparing ciphertext with a constant, as the encryption function is only additively homomorphic. Although there are methods of achieving this operation by subtracting the comparison constant $c$ from the encrypted number $enc(X)$, then multiplying the difference with a randomly generated positive number $r$, sending the final value $\left(enc(X) - c\right) \times r$ to be decrypted using the appropriate private key, and seeing whether the result is a positive or negative number, the added complexity and communication cost make this implementation undesirable for PPBSP. Since the trading platform will only ever need to compare the individual deviations with 0, checking if they are positive or negative, the SM can perform this quick operation itself, as it already knows the comparison constant. Therefore, a simple binary value sent by the user's SM inside the payload is enough.
          \end{itemize}
          \item It encrypts the privacy-sensitive data using partially homomorphic encryption:
          \begin{itemize}
            \item $U^{x}_{P2P}$ is encrypted once using the supplier's homomorphic public key $pub\_S_k$, and once using the grid operator's analogous public key $pub\_GridOp$, generating two separate ciphertexts: $\{U^{x}_{P2P}\}_{pub\_S_k}$ and $\{U^{x}_{P2P}\}_{pub\_GridOp}$
            \item $InDev_x$ is similarly encrypted once using the supplier's homomorphic public key $pub\_S_k$, and once using the grid operator's key $pub\_GridOp$, generating another two separate ciphertexts: $\{InDev_x\}_{pub\_S_k}$ and $\{InDev_x\}_{pub\_GridOp}$
            \item Since $is\_bid\_accepted$, $bid\_type$, $net\_consumption\_type$, and $sign(InDev_x)$ contain less detailed private user information, their privacy-invasiveness must be weighed up against their necessity in the billing algorithm. Therefore, they are not homomorphically encrypted in order to support the comparison operation performed by the trading platform, which is not natively supported by the Paillier cryptosystem, despite the small, but non-zero, information leak.
          \end{itemize}
          If PPBSP were to perform homomorphic encryption only using the user's corresponding supplier's public key $pub\_S_k$, bill settlements would still be possible for the first two billing models (see Sections~\ref{sec:billing_1} and~\ref{sec:billing_2}), but market-wide energy consumption deviation data across multiple suppliers' customer bases could no longer be aggregated, rendering the last two billing algorithms unusable (see Sections~\ref{sec:billing_3} and~\ref{sec:billing_4}). Moreover, by not rerunning the calculations with an independent trusted third-party (TTP) homomorphic public key, we would not be able to fulfil one of this project's proposed requirements, leaving no entity capable of holding the suppliers accountable by potentially verifying their bill settlement results. \\ \ \\
          Alternatively, the user's consumption data could only be encrypted using the grid operator's key $pub\_GridOp$, which still allows for any of the four possible partial bill calculation algorithms to be used by the trading platform. However, in this case, the computation and communication load placed on the GridOp would be disproportionately large, and it would also introduce an unnecessary single point of failure (SPOF) into the system, leaving the GridOp as the only entity capable of decrypting and reporting each user's bills and each supplier's profits. \\ \ \\
          Therefore, the payload includes multiple encrypted versions of the consumption data, once with the corresponding supplier's public key $pub\_S_k$, and once with the grid operator's public key $pub\_GridOp$, increasing the redundancy of the system by effectively doubling the encryption computation load on the SM and the partial bill calculation time on the trading platform, neither of which are too high anyway, but crucially eliminating the SPOF, thus improving the system's availability and reliability, and also minimising any risk of a supplier deviating from the protocol, because of the implicit auditing ability of the grid operator.
          \item It appends the aforementioned data values together to construct the payload and send it to the trading platform:
          \begin{equation}
          \begin{aligned}
                payload = (
                & is\_bid\_accepted \| bid\_type \| net\_consumption\_type \| sign(InDev_x) \| \\
                & \{U^x_{P2P}\}_{pub\_S_k} \|\{InDev_x\}_{pub\_S_k} \| \{U^{x}_{P2P}\}_{pub\_GridOp} \|   \{InDev_x\}_{pub\_GridOp} )
            \end{aligned}
            \end{equation}
            Naturally, the entire payload is further encrypted in line with the public-key cryptosystem used to implement the underlying distributed network of computers. The communication protocol between the network's devices is assumed to be secure, protected against tempering by external entities, and thus has been abstracted from this and all upcoming steps of the bill settlement process.
        \end{enumerate}
    
    \subsection{Trading Platform Protocol}
            The responsibility of the trading platform (TrPlat) is two-fold:
            \begin{enumerate}
                \item Calculate the partial bills of the market's users.
                \item Communicate the final bills to the corresponding suppliers.
            \end{enumerate}
        Firstly, every trading period (e.g. 30 minutes, one hour, etc.), the trading platform receives the payloads from its users (e.g. households, small businesses, etc.) which include homomorphically encrypted versions of their fine-grained energy consumption data, along with information regarding their bid/offer at the latest electricity auction. These payloads are individually stored by the TrPlat, while the individual deviation data which is encrypted using the grid operator's homomorphic public key $pub\_GridOp$ is aggregated into market-wide statistics that represent the deviations of energy supply and demand from their predicted values, described in detail in Section~\ref{sec:gridop}. The encrypted aggregate data is then sent to the GridOp for decryption. Depending on the selected billing model implemented by the trading platform, the TrPlat could continue with the partial bill calculation immediately after receiving an acknowledgement message from the GridOp (see Sections~\ref{sec:billing_1} and~\ref{sec:billing_2}), or it might have to wait for the plaintext version of the aggregated statistics to arrive back from the GridOp before being able to pursue its algorithm any further, as the decrypted values are vital parts of the bill calculation algorithm (see Sections~\ref{sec:billing_3} and~\ref{sec:billing_4}). After the trading slot's partial bills are computed, the TrPlat sends each supplier its own respective net balance change for that period, representing the sum of all buying/selling energy transactions with its users at the retail market (positive for making a profit, negative for making a loss), in order to help them better understand the direction the market is leaning towards (demand or supply), leading to more accurate predictions of future customer behaviour trends and making them better prepared to meet their customers' demands in the following trading periods.\\ \ \\
        Secondly, every billing period (e.g. one month), the trading platform needs to inform the suppliers of their respective customers' individual final energy bills, homomorphically encrypted using the specific supplier's public key $pub\_S_k$, which are more likely to be negative (owing money to their supplier), than positive (being owed money by their supplier). In most of the billing models presented in this report (see Sections~\ref{sec:billing_2},~\ref{sec:billing_3}, and~\ref{sec:billing_4}), these final bills include both the bills to/from the P2P market (from energy transactions with other users) and the bills to/from the RM (from transactions directly with the energy supplier), in a single numerical value which each user pays to/receives from their supplier. The details of bill settlement and supplier responsibilities are described in Section~\ref{sec:supplier}. 
        \clearpage
     
        \subsubsection{Overview of Billing Models} \label{sec:billing}
    The following subsections outline different billing algorithms for P2P markets with imperfect bid-offer fulfillment which run after every trading period on the trading platform, whose inputs include each user's energy volume committed to the P2P market and their individual deviation from that value, and whose outputs consist of the users' partial bills and each supplier's balance change for the respective trading slot. Evidently, all input and output values are homomorphically encrypted ciphertext, in order to preserve the privacy of user bills and fine-grained energy measurements. Moreover, each billing algorithm is run twice every trading period, once using inputs encrypted with the corresponding supplier's public key $pub\_S_k$ and once using those encrypted with the grid operator's public key $pub\_GridOp$. The rationale behind this requirement is comprehensively explained in step 2 of Section~\ref{sec:user}.
    \\ \ \\
    The four implemented billing models have been adapted from~\cite{madhusudan2022billing}, with all of them being modified to allow for the use of Paillier homomorphic encryption. Furthermore, the Status Quo algorithm (see Section~\ref{sec:billing_1}) has been adapted to fit the type of payload described in Section~\ref{sec:user}, and the Social Cost Split (see Section~\ref{sec:billing_3}) and Universal Cost Split (see Section~\ref{sec:billing_4}) billing models have also been improved by adding a weighted redistribution of the total volume of energy over-consumed (under-produced) or under-consumed (over-produced) by reselling it at the trading price (TP) to each user whose $InDev_x$ negatively influenced the balance of the electricity grid (pushed the corresponding aggregated deviation value away from 0), according to each respective user's proportional contribution to the total deviation TD, total supply deviation TSD, and total demand deviation TDD, instead of the equal redistribution implemented and analysed by~\cite{madhusudan2022billing}. Notations are listed in the \hyperlink{table:not}{Notations section}.
    \begin{figure}[h]
              \centering
              \includegraphics[width=1.0\textwidth]{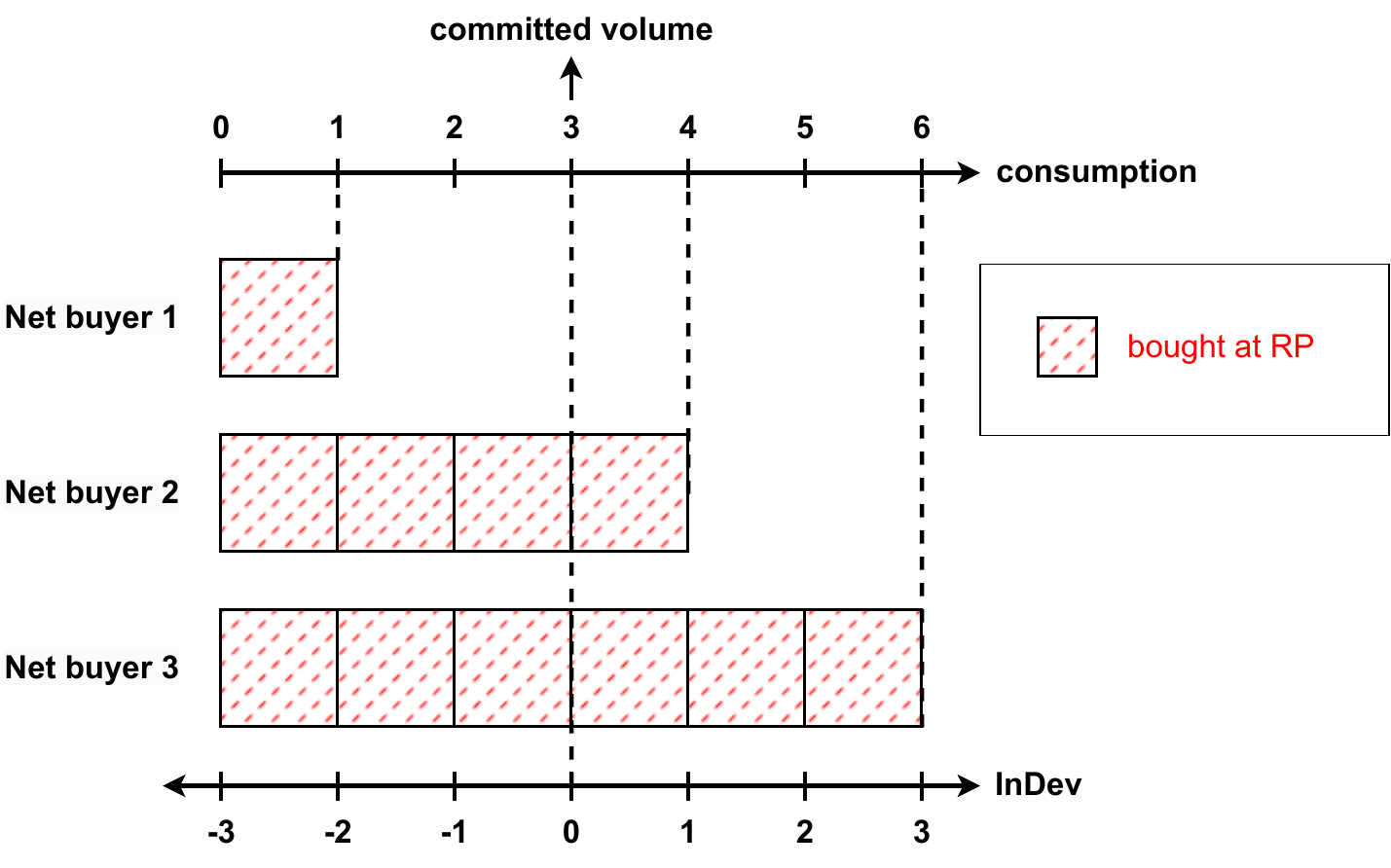}
              \caption{Status quo billing example.}
              \label{fig:billing1}
    \end{figure}
        \subsubsection{Billing Model for Retail Markets – the Status Quo} \label{sec:billing_1}
    The status quo in most of the liberalised energy retail markets (RM) around the world allows consumers and consumers to trade electrical energy only directly with their contracted supplier at the RM. This exclusivity deal forbids any energy trading between local consumers and prosumers, their only option being buying to/selling from one of the accredited energy providers. Any excess of energy produced by prosumers that is fed into the power grid is regulated and bought by suppliers at a unique tariff, called the Feed-In Tariff (FiT), which is standardised at the country level by the respective market regulator. Therefore, users have little to no incentive to adopt renewable energy sources or change their load profiles. Moreover, in the presence of an imperfect bill-offer fulfillment assumption, whereby the users' actual consumption data is expected to deviate slightly from the predicted values, neither consumers nor prosumers have any reason to reduce their individual deviations, which would help minimise the grid-balancing cost. \\ \ \\
    The billing model that illustrates the status quo has been described in detail in Algorithm~\ref{algo:retail_market}. The version of the algorithm presented in this report has been designed to function for both Non-P2P and P2P users, using the payload format outlined in Section~\ref{sec:user}. In the case of Non-P2P users, it is just a privacy-preserving version of the worldwide status quo, while for P2P users, it represents the fall-back option for bids that were not accepted at the P2P market, which still need to be settled somehow. However, the desire is for as many bids as possible to be accepted at the energy auction, such that fewer partial bills are calculated using this billing model. \\ \ \\
    The appropriate naming convention for the two types of users in the RM billing model is net buyers and net sellers, with the former consuming more energy than they produce, and the latter doing the opposite. This is a departure from the usual nomenclature used throughout this report, which otherwise splits users into consumers and prosumers, because these terms only refer to a user's type of bid/offer at the energy auction, rather than their actual meter readings with which the Status Quo algorithm is concerned. Figure~\ref{fig:billing1} showcases an example market with three net buyers, whose entire need for electricity is fulfilled at the RM. \\ \ \\
    A summary of the privacy-enhancing status quo algorithm is given below:
    \begin{algorithm}[t]
        \caption{Billing Model for Retail Markets}\label{algo:retail_market}
        \begin{algorithmic}[1]
            \Procedure{Net buyer bills, Net seller rewards, supplier balance}{}
            \For{each timeslot}
            \For{each $x$, $k$ in $U_n$, $S_n$}
                \If{user is P2P}
                \State $\{U^x_{val}\}_{pub\_S_k}$ = $\{U^{x}_{P2P}\}_{pub\_S_k}$ + $\{InDev_x\}_{pub\_S_k}$
                \EndIf
                \If{$net\_consumption\_type$ $\not=$ $bid\_type$}
                \State $\{U^x_{val}\}_{pub\_S_k}$ = $- \{U^x_{val}\}_{pub\_S_k}$
                \EndIf
                \If{user is net buyer}
                \State $\{U^x$ bill$\}_{pub\_S_k}$ = $\{U^x_{val}\}_{pub\_S_k}$ $\times$ RP 
                \State $\{S^{inc}_k\}_{pub\_S_k}$ += $\{U^x$ bill$\}_{pub\_S_k}$
                \EndIf
                \If{user is net seller}
                \State $\{U^x$ reward$\}_{pub\_S_k}$ = $\{U^x_{val}\}_{pub\_S_k}$ $\times$ FiT 
                \State $\{S^{exp}_k\}_{pub\_S_k}$ += $\{U^x$ reward$\}_{pub\_S_k}$
                \EndIf
                \State $\{S^{bal}_k\}_{pub\_S_k}$ += $\{S^{inc}_k\}_{pub\_S_k} - \{S^{exp}_k\}_{pub\_S_k}$
            \EndFor
            \EndFor
            \EndProcedure
        \end{algorithmic}
    \end{algorithm}
    \begin{itemize}
        \item The payloads of P2P users, which separate the half-hourly consumption data into the volume committed to the P2P market $\{U^{x}_{P2P}\}_{pub\_S_k}$ and the household's deviation from that volume $\{InDev_x\}_{pub\_S_k}$, are converted back into a single value for the actual meter reading $\{U^x_{val}\}_{pub\_S_k}$. The encrypted consumption data in Non-P2P payloads is already formatted in the appropriate style and does not require any alterations.
        \item Net buyers can purchase electricity solely from their suppliers at a retail price (RP). The suppliers set the RP, and typically present consumers with multiple tariffs to choose from, ensuring that their prices remain competitive with those of other energy suppliers.
        \item Net sellers are limited to selling their surplus of electrical energy to suppliers at a fixed rate, commonly referred to as FiT. Depending on the country, the FiT may vary based on the type and size of the renewable energy source used by the user to generate electricity.
        \item Energy suppliers are responsible for selling electricity to the customers with whom they have signed exclusive contracts, while also buying back any excess electricity that these users return to the grid.
    \end{itemize}
    With all electricity transactions involving the suppliers, the volume traded at the RM for the status quo is equal to the sum of all measured consumption volumes:
        \begin{equation}
                V^{RM}_{SQ} = \sum_{x=1}^{U_n}|U^x_{val}|
        \end{equation}
    It must be noted that suppliers are very likely to profit from transactions at the RM, especially when generalising to the level of their entire customer base, since the price they sell energy at, the RP, is much higher than the price at which they buy energy back from the users, the FiT. For example, even in a case where the energy consumption of the user base is equal to its surplus energy production, and thus the volume sold is equal to the volume bought at the RM, the supplier, being the only allowed trading partner for any individual customer, will always accrue a sizeable profit. Therefore, the status quo greatly favours the suppliers, giving little-to-no incentive for net sellers to ever produce more electrical energy than they themselves require, as the excess volume sold at the RM will only grant them a marginal capital gain. \\ \ \\
    With this in mind, the need for billing models for P2P markets that allow users to trade energy among themselves becomes apparent. Three such models for P2P markets with the RM as back-up are described in the following subsections.
    \begin{figure}[h]
              \centering
              \includegraphics[width=1.0\textwidth]{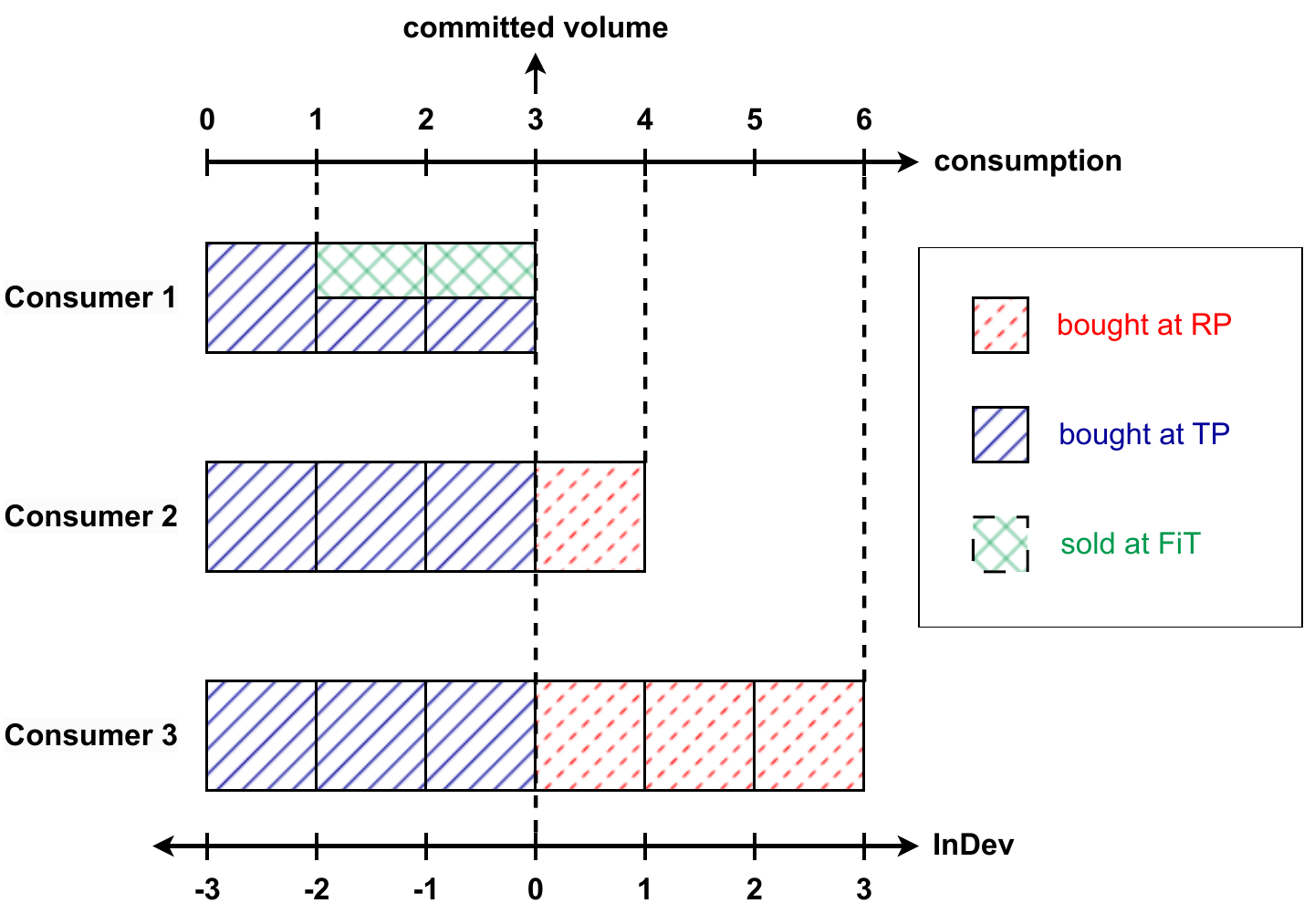}
              \caption{Individual cost split billing example.}
              \label{fig:billing2}
    \end{figure}
        \subsubsection{Billing Model with Individual Cost Split} \label{sec:billing_2}
    This billing algorithm makes each P2P user independently responsible for trading away/compensating for their individual deviations from the committed bids/offers, by buying/selling the electricity deficit/surplus at the retail market (RM), while the committed volumes are traded at trading price (TP). Negative deviations must be compensated for at the RM in order for the P2P trade commitments to be fulfilled, while positive deviations of both consumers and prosumers are also traded directly at this back-up market. Figure~\ref{fig:billing2} illustrates an example market with three individual consumers whose committed volumes are all 3 kWh, highlighting the use of the RM to either sell their negative deviation (Consumer 1) or purchase their positive deviations (Consumer 2 and Consumer 3). \\ \ \\
    The specifics of the privacy-preserving Individual Cost Split billing model are shown in Algorithm~\ref{algo:ind_split} and are outlined below:
    \begin{algorithm}[t]
                \caption{Billing Model with Individual Cost Split}\label{algo:ind_split}
                \begin{algorithmic}[1]
                \Procedure{Customer bills, Prosumer rewards, supplier balance}{}
                \For{each timeslot}
                \If{bid accepted}
                \For{each $i$, $j$ and $k$ in $P2P^c_n$, $P2P^p_n$ and $S_n$}
                \If{$sign(InDev_x) = 0$}
                \State $\{P2P^i_c$ bill$\}_{pub\_S_k}$ = $\{C^{P2P}_{dem}\}_{pub\_S_k}$ $\times$ TP
                \State $\{P2P^j_p$ reward$\}_{pub\_S_k}$ = $\{P^{P2P}_{sup}\}_{pub\_S_k}$ $\times$ TP
                \State $\{S^{inc}_k\}_{pub\_S_k}$ += 0; $\{S^{exp}_k\}_{pub\_S_k}$ += 0
                \EndIf
                \If{$sign(InDev_x) < 0$}
                \State $\{P2P^i_c$ bill$\}_{pub\_S_k}$ = $\{C^{P2P}_{dem}\}_{pub\_S_k}$ $\times$ TP + $\{InDev_i\}_{pub\_S_k}$ $\times$ FiT
                \State $\{P2P^j_p$ reward$\}_{pub\_S_k}$ = $\{P^{P2P}_{sup}\}_{pub\_S_k}$ $\times$ TP + $\{InDev_j\}_{pub\_S_k}$ $\times$ RP   
                \State $\{S^{inc}_k\}_{pub\_S_k}$ -= $\{InDev_j\}_{pub\_S_k}$ $\times$ RP
                \State $\{S^{exp}_k\}_{pub\_S_k}$ -= $\{InDev_i\}_{pub\_S_k}$ $\times$ FiT
                \EndIf
                \If{$sign(InDev_x) >0$}
                \State $\{P2P^i_c$ bill$\}_{pub\_S_k}$ = $\{C^{P2P}_{dem}\}_{pub\_S_k}$ $\times$ TP + $\{InDev_i\}_{pub\_S_k}$ $\times$ RP
                \State $\{P2P^j_p$ reward$\}_{pub\_S_k}$ = $\{P^{P2P}_{sup}\}_{pub\_S_k}$ $\times$ TP + $\{InDev_j\}_{pub\_S_k}$ $\times$ FiT
                \State $\{S^{inc}_k\}_{pub\_S_k}$ += $\{InDev_i\}_{pub\_S_k}$ $\times$ RP
                \State $\{S^{exp}_k\}_{pub\_S_k}$ += $\{InDev_j\}_{pub\_S_k}$ $\times$ FiT
                \EndIf
                \State $\{S^{bal}_k\}_{pub\_S_k}$ = $\{S^{inc}_k\}_{pub\_S_k} - \{S^{exp}_k\}_{pub\_S_k}$
                \EndFor
                \EndIf
                \If{bid not accepted}
                \For{each $i$, $j$ and $k$ in $C_n$, $P_n$ and $S_n$}
                \State \textbf{goto} Algorithm~\ref{algo:retail_market}
                \EndFor
                \EndIf
                \EndFor
                \EndProcedure
                \end{algorithmic}
        \end{algorithm}
    \begin{itemize}
        \item Positive deviations of consumers (i.e when a consumer consumes more energy than their committed volume) are bought at RP from their contracted energy supplier.
        \item Negative deviations of consumers (i.e. when a consumer consumes less energy than they committed in their bid) are also compensated at the RM. The committed energy volume must still be purchased in its entirety from the P2P market at TP as the bid dictates, despite not being consumed completely, and is then sold at a fixed FiT to their corresponding energy supplier.
        \item Positive deviations of prosumers (i.e when a prosumer produces more electricity than their committed volume) are sold directly to their supplier at the FiT.
        \item Negative deviations of prosumers (i.e. when a prosumer produces less electricity than they committed in their offer) are also compensated at the RM. Since the prosumer must nevertheless offer to the P2P market the volume of energy they committed, the difference is bought from their contracted supplier at RP and immediately sold at TP to the P2P market.
        \item Consumers/prosumers with no individual deviation only trade their energy volumes at TP.
        \item Energy suppliers are responsible for selling electricity at RP to under-supplying prosumers ($InDev_j < 0$) and over-consuming consumers ($InDev_i > 0$). They also purchase electricity at FiT from over-supplying prosumers ($InDev_j > 0$) and under-consuming consumers ($InDev_i < 0$).
        \item Consumers/prosumers whose bids/offers were not accepted at the auction trade their entire consumption/production with the respective supplier at RP and FiT, according to Algorithm~\ref{algo:retail_market}.
    \end{itemize}
    With all individual deviations being independently traded at the RM, whether bought or sold, the total volume of energy traded with the suppliers is equal to the sum of the absolute values of every consumer's/prosumer's deviation:
        \begin{equation}
                V^{RM}_{Ind} = \sum_{i=1}^{P2P^c_n}|InDev_i| + \sum_{j=1}^{P2P^p_n}|InDev_j|
        \end{equation}
    However, it is not ideal for individual deviations to be traded independently of each other. For example, assuming a system with 2 consumers, the first having a deviation $InDev_1 = 2$, and the other $InDev_2 = -2$, there is effectively no need for each of them to resolve their deviations with their supplier independently, as they would both benefit from compensating for one another, since an under-consumer balances out an equivalent over-consumer on the power grid. Intuitively, the under-consumer sells the unused electricity they bought at TP from the P2P market to the over-consumer at the same TP. The former benefits from selling their excess energy at TP instead of the lower FiT, while the latter benefits from buying their electricity deficit at TP instead of the higher RP. Therefore, in order to minimise the volume traded at the RM, reduce consumer bills and increase prosumer rewards, it is possible for the consumers (over-consuming and under-consuming) to compensate for each other, with the prosumers (over-supplying and under-supplying) also doing the same. Such a billing model is introduced in the next subsection.
    \begin{figure}[h]
              \centering
              \includegraphics[width=1.0\textwidth]{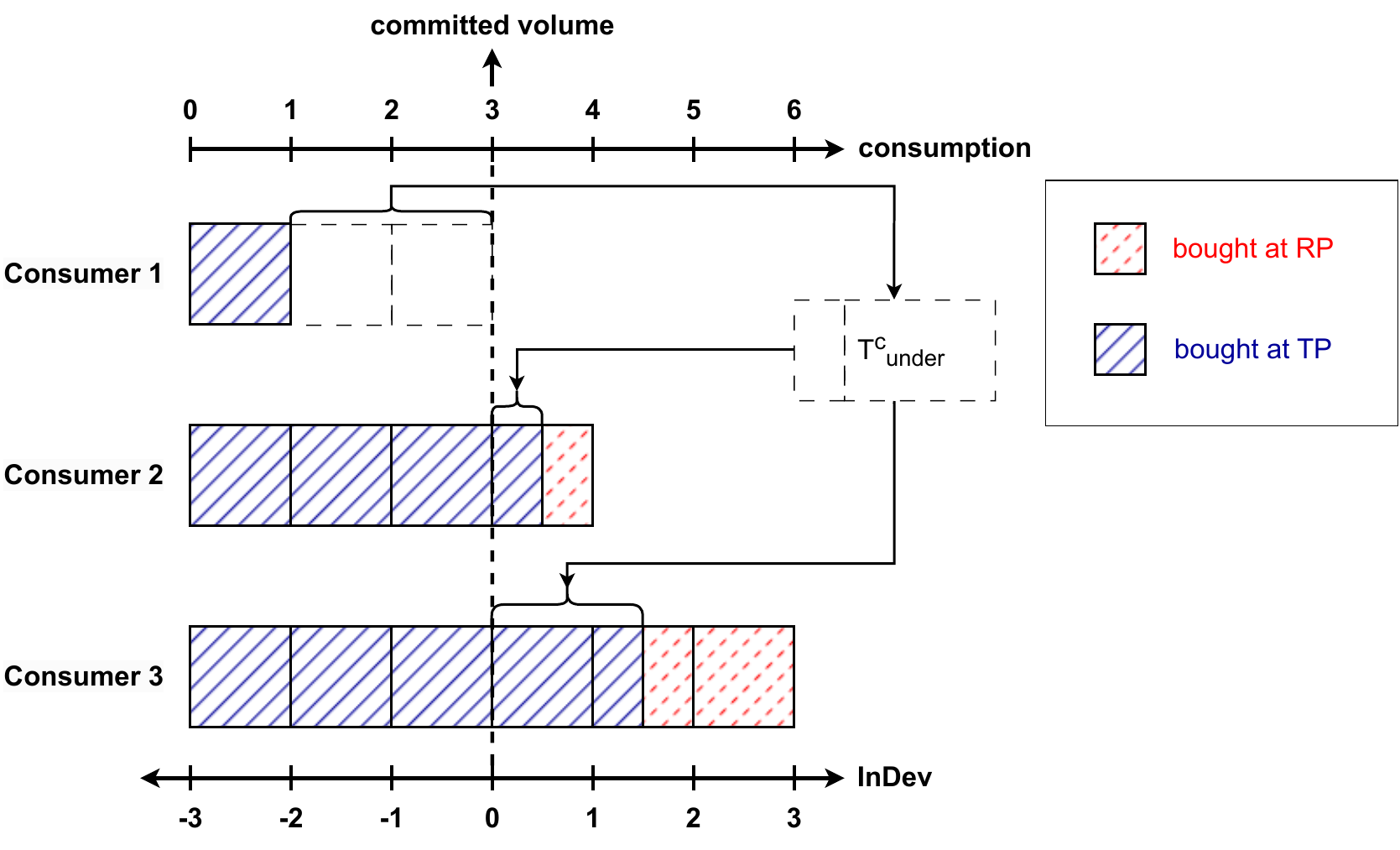}
              \caption{Weighted social cost split billing example.}
              \label{fig:billing3}
    \end{figure}
        \subsubsection{Billing Model with Weighted Social Cost Split} \label{sec:billing_3}
    This billing model uses the aggregated individual deviations of consumers to socially split the cost among them proportionally to each consumer household's effect on the total demand deviation. Similarly, the prosumers' deviations are aggregated into a single value which is split among prosumers according to their contribution to the total supply deviation. Because this model makes use of aggregated values of encrypted meter readings in calculating the partial bills, the trading platform must wait for the decrypted statistics to arrive from the grid operator before continuing its algorithm. Figure~\ref{fig:billing3} illustrates an over-consuming market with three individual consumers whose committed volumes are all 3 kWh, highlighting the weighted redistribution of electricity bought at TP from the one under-consumer (Consumer 1) to the two over-consumers (Consumer 2 and Consumer 3), before using the RM as a fall-back. \\ \ \\
    The characteristics of the privacy-preserving Weighted Social Cost Split billing model are shown in Algorithm~\ref{algo:social_split} and described below:
    \begin{algorithm}
                \caption{Billing Model with Weighted Social Cost Split}\label{algo:social_split}
                \begin{algorithmic}[1]
                \scriptsize
                \For{each timeslot}
                \If{bid accepted}
                \Procedure{Customer bills, supplier income/expenditure}{}
                \For{each $i$ and $k$ in $P2P^c_n$ and $S_n$}
                \If{TDD = 0}
                \State $\{P2P^i_c$ bill$\}_{pub\_S_k}$ = $\{C^{P2P}_{dem}\}_{pub\_S_k}$ $\times$ TP
                \State $\{S^{inc}_k\}_{pub\_S_k}$ += 0
                \EndIf
                \If{TDD $< 0$}
                \If{$sign(InDev_i) \geq 0$}
                \State $\{P2P^i_c$ bill $\}_{pub\_S_k}$ = $(\{C^{P2P}_{dem}\}_{pub\_S_k} + \{InDev_i\}_{pub\_S_k})$ $\times$ TP
                \State $\{S^{inc}_k\}_{pub\_S_k}$ += 0
                \EndIf
                \If{$sign(InDev_i) < 0$}
                \State $\{P2P^i_c$ bill$\}_{pub\_S_k}$ = $(\{C^{P2P}_{dem}\}_{pub\_S_k} + \{InDev_i\}_{pub\_S_k} \times \frac{T^c_{over}}{T^c_{under}}) \times$ TP $ + \{InDev_i\}_{pub\_S_k} \times (1 - \frac{T^c_{over}}{T^c_{under}}) \times$ FiT
                \State $\{S^{exp}_k\}_{pub\_S_k}$ -= $\{InDev_i\}_{pub\_S_k} \times (1 - \frac{T^c_{over}}{T^c_{under}}) \times$ FiT
                \EndIf
                \EndIf
                \If{TDD $> 0$}
                \If{$sign(InDev_i) \leq 0$}
                \State $\{P2P^i_c$ bill$\}_{pub\_S_k}$ = $(\{C^{P2P}_{dem}\}_{pub\_S_k} + \{InDev_i\}_{pub\_S_k})$ $\times$ TP
                \State $\{S^{inc}_k\}_{pub\_S_k}$ += 0
                \EndIf
                \If{$sign(InDev_i) > 0$}
                \State $\{P2P^i_c$ bill$\}_{pub\_S_k}$ = $(\{C^{P2P}_{dem}\}_{pub\_S_k} + \{InDev_i\}_{pub\_S_k} \times \frac{T^c_{under}}{T^c_{over}}) \times$ TP $ + \{InDev_i\}_{pub\_S_k} \times (1 - \frac{T^c_{under}}{T^c_{over}}) \times$ RP
                \State $\{S^{inc}_k\}_{pub\_S_k}$ += $\{InDev_i\}_{pub\_S_k} \times (1 - \frac{T^c_{under}}{T^c_{over}}) \times$ RP
                \EndIf
                \EndIf
                \State $\{S^{bal}_k\}_{pub\_S_k}$ = $\{S^{inc}_k\}_{pub\_S_k} - \{S^{exp}_k\}_{pub\_S_k}$
                \EndFor
                \EndProcedure
                \Procedure{Prosumer rewards, supplier income/expenditure}{}
                \For{each $j$ and $k$ in $P2P^p_n$ and $S_n$}
                \If{TSD = 0}
                \State $\{P2P^j_p$ reward$\}_{pub\_S_k}$ = $\{P^{P2P}_{sup}\}_{pub\_S_k}$ $\times$ TP
                \State $\{S^{exp}_k\}_{pub\_S_k}$ += 0
                \EndIf
                \If{TSD $< 0$}
                \If{$sign(InDev_j) \geq 0$}
                \State $\{P2P^j_p$ reward$\}_{pub\_S_k}$ = $(\{P^{P2P}_{sup}\}_{pub\_S_k} + \{InDev_j\}_{pub\_S_k})$ $\times$ TP
                \State $\{S^{exp}_k\}_{pub\_S_k}$ += 0
                \EndIf
                \If{$sign(InDev_j) < 0$}
                \State $\{P2P^j_p$ reward$\}_{pub\_S_k}$ = $(\{P^{P2P}_{sup}\}_{pub\_S_k} + \{InDev_j\}_{pub\_S_k} \times \frac{T^p_{over}}{T^p_{under}}) \times$ TP $ + \{InDev_j\}_{pub\_S_k} \times (1 - \frac{T^p_{over}}{T^p_{under}}) \times$RP
                \State $\{S^{inc}_k\}_{pub\_S_k}$ -= $\{InDev_j\}_{pub\_S_k} \times (1 - \frac{T^p_{over}}{T^p_{under}}) \times$ RP
                \EndIf
                \EndIf
                \If{TSD $> 0$}
                \If{$sign(InDev_j) \leq 0$}
                \State $\{P2P^j_p$ reward$\}_{pub\_S_k}$ = $(\{P^{P2P}_{sup}\}_{pub\_S_k} + \{InDev_j\}_{pub\_S_k})$ $\times$ TP
                \State $\{S^{exp}_k\}_{pub\_S_k}$ += 0
                \EndIf
                \If{$sign(InDev_j) > 0$}
                \State $\{P2P^j_p$ reward$\}_{pub\_S_k}$ = $(\{P^{P2P}_{sup}\}_{pub\_S_k} + \{InDev_j\}_{pub\_S_k} \times \frac{T^p_{under}}{T^p_{over}}) \times$ TP $ + \{InDev_j\}_{pub\_S_k} \times (1 - \frac{T^p_{under}}{T^p_{over}}) \times$FiT
                \State $\{S^{exp}_k\}_{pub\_S_k}$ += $\{InDev_j\}_{pub\_S_k} \times (1 - \frac{T^p_{under}}{T^p_{over}}) \times$ FiT
                \EndIf
                \EndIf
                \State $\{S^{bal}_k\}_{pub\_S_k}$ = $\{S^{inc}_k\}_{pub\_S_k} - \{S^{exp}_k\}_{pub\_S_k}$
                \EndFor
                \EndProcedure
                \EndIf
                \If{bid not accepted}
                \For{each $i$, $j$ and $k$ in $C_n$, $P_n$ and $S_n$}
                \State \textbf{goto} Algorithm~\ref{algo:retail_market}
                \EndFor
                \EndIf
                \EndFor
                \end{algorithmic}
        \end{algorithm}
    \begin{itemize}
        \item The total demand deviation (TDD) is calculated as the sum of all consumer individual deviations. The total supply deviation (TSD) represents the aggregate of all prosumer individual deviations.
        \item If the TDD is equal to zero, then all consumers buy their entire consumed energy volume at TP, regardless of their individual deviations. This benefits both under-consumers, since they do not need to sell their excess electricity at FiT to the RM, and the over-consumers, as they are not forced to buy their electricity deficit at RP from their supplier. The case where the TSD is zero is analogous. 
        \item A positive TDD indicates that the consumers, as a whole, over-consumed in relation to the total volume bought at the P2P market. Therefore, the under-consumers only partially compensate for the over-consumers. Effectively, the under-consumers buy their committed energy volumes at TP and then sell on the unused electricity to the over-consumers also at TP. Such, the total volume under-consumed $T^c_{under}$ disappears from the under-consumers' bills and is proportionately redistributed to the over-consumers' bills, based on each over-consumer's contribution to the total volume over-consumed $T^c_{over}$. In practice, consumers that under-consumed buy their actual energy consumption volume at TP, regardless of their individual deviation, while the consumers who over-consumed buy their committed volume at TP, an additional proportion of the compensated energy at TP, and the rest of their consumption at RP from the RM. As a result, under-consumers benefit from buying their exact energy consumption at TP, rather than buying the larger committed volume at TP and selling the deviation at FiT, while the over-consumers gain by buying part of their individual deviation at TP and the rest at RP, instead of purchasing it all at RP.  
        \item A negative TDD indicates that the consumers, as a whole, under-consumed, leaving excess energy on the grid, and the over-consumers can only partly compensate for the under-consumers. In essence, the under-consumers again buy their committed energy volumes at TP and then sell on proportions of their individual deviations to the over-consumers also at TP. The proportions resold at TP are based on each under-consumer's contribution to the total volume under-consumed $T^c_{under}$ and add up to the value of the total volume over-consumed $T^c_{over}$. Therefore, the over-consumers purchase their entire actual energy consumption volume at TP, whereas the under-consumers, after reselling an appropriate part of their individual deviation at TP, sell the rest at FiT to the supplier. As a result, the over-consumers benefit by buying their total energy consumption at TP, instead of buying the committed volume at TP and their deviation at RP, and the under-consumers gain by selling part of their individual deviation at TP and the rest at FiT, rather than selling it all at FiT.
        \item Similarly, the cases for prosumers are analogous. If the TSD is positive (negative), then the aggregate deviation of prosumers who under-supplied (over-supplied) partially compensates for the prosumers who over-supplied (under-supplied). Under-suppliers (over-suppliers) sell all their produced electricity at the TP regardless of their individual deviation, while the prosumers who over-supplied (under-supplied) reduce their individual deviation with a proportional share of the total deviation of consumers who under-supplied $T^p_{under}$ (over-supplied $T^p_{over}$), determined by their contribution to the total volume over-supplied $T^p_{over}$ (under-supplied $T^p_{under}$). Therefore, prosumers benefit either by selling their entire production at TP or by reducing the revenue loss incurred by their individual deviations, which must be compensated for at the RM.
        \item Energy suppliers only trade electricity with those consumers/prosumers whose individual deviation sign is the same as the sign of the TDD/TSD respectively.
        \item Users whose bids/offers were not accepted at the auction must trade their entire consumption/production at the retail market at RP and FiT, according to Algorithm~\ref{algo:retail_market}.
    \end{itemize}
    By dealing with consumer bills separately from prosumer rewards, the total volume of electricity traded with the energy suppliers is equal to the sum of the absolute values of TDD and TSD, with the former representing the sum of all consumer deviations, and the latter constituting the sum of all prosumer deviations:
        \begin{equation}
                V^{RM}_{Soc} = |\sum_{i=1}^{P2P^c_n}InDev_i| + |\sum_{j=1}^{P2P^p_n}InDev_j|
        \end{equation}
    Clearly, the billing model is still not optimal because of this separation between the consumers' and prosumers' effect on the balance of the grid. For instance, assuming a total supply deviation TSD $= 2$ and a total demand deviation TDD $= 2$, both consumers and prosumers would benefit from compensating for one another. Intuitively, the users which increase the total deviation of the grid (TD) by producing a surplus of electricity (over-suppliers) or by not fulfilling their consumption commitments (under-consumers) sell their individual deviations at the TP to the households which decrease the TD by producing less energy than they committed to (under-suppliers) or by consuming more electricity than they bid for (over-consumers). It is also important to note that a positive value of TSD has the same effect on the load on the power grid as an opposite-valued negative TDD, with the former indicating an over-consumption trend, and the latter an under-production tendency, both leading to a deficit of electricity. The opposite also holds. A billing algorithm that implements this idea is presented next.
    \begin{figure}[h]
              \centering
              \includegraphics[width=1.0\textwidth]{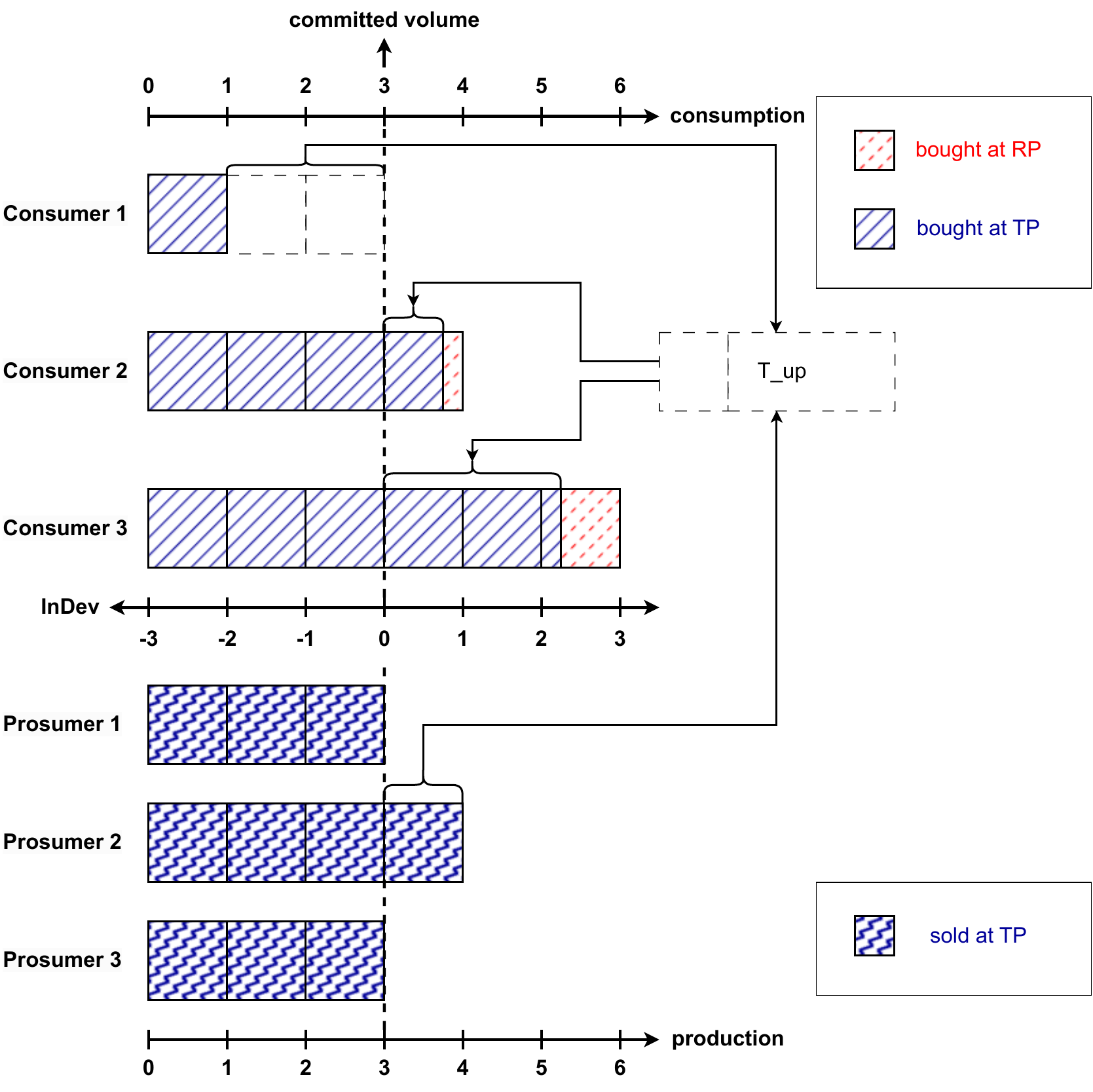}
              \caption{Weighted universal cost split billing example.}
              \label{fig:billing4}
    \end{figure}
        \subsubsection{Billing Model with Weighted Universal Cost Split} \label{sec:billing_4}
    The final billing model is concerned with the total deviation of the P2P market, an aggregate value of all individual deviations of both consumers and prosumers, which is split among those users whose deviations are in the same direction as the TD, in proportion to their contribution to the aggregate value. Similarly to the Weighted Social Cost Split (see Section~\ref{algo:social_split}), the trading platform is forced to wait for the communication of the decrypted aggregate P2P market statistics before starting the partial bill calculation phase.  Figure~\ref{fig:billing4} showcases an example market with three individual consumers and three individual prosumers whose committed volumes are all 3 kWh, highlighting the weighted redistribution of electricity at TP from the one under-consumer (Consumer 1) and one over-supplier (Prosumer 2) to the two over-consumers (Consumer 2 and Consumer 3), before using the RM as a fall-back. \\ \ \\
    For the rest of this section, users who push the TD up by under-consuming or over-supplying will also be referred to as uptrenders, while those who pull the TD down by over-consuming or under-supplying will also be called downtrenders. The exact implementation of the privacy-preserving Weighted Universal Cost Split billing model is illustrated in Algorithm~\ref{algo:univ_split} and explained below:
    \begin{algorithm}
                \caption{Billing Model with Weighted Universal Cost Split}\label{algo:univ_split}
                \begin{algorithmic}[1]
                \scriptsize
                \Procedure{Customer bills, Prosumer rewards, supplier balance}{}
                \For{each timeslot}
                \If{bid Accepted}
                \For{each $i$, $j$ and $k$ in $P2P^c_n$, $P2P^p_n$ and $S_n$}
                \If{TD = 0}
                \State $\{P2P^i_c$ bill$\}_{pub\_S_k}$ = $\{C^{P2P}_{dem}\}_{pub\_S_k}$ $\times$ TP
                \State $\{P2P^j_p$ reward$\}_{pub\_S_k}$ = $\{P^{P2P}_{sup}\}_{pub\_S_k}$ $\times$ TP
                \State $\{S^{inc}_k\}_{pub\_S_k}$ += 0; $\{S^{exp}_k\}_{pub\_S_k}$ += 0
                \EndIf
                \If{TD $<$ 0}
                \State $\{S^{exp}_k\}_{pub\_S_k}$ += 0
                \If{$sign(InDev_x) \leq 0$}
                \State $\{P2P^i_c$ bill$\}_{pub\_S_k}$ = $(\{C^{P2P}_{dem}\}_{pub\_S_k} + \{InDev_i\}_{pub\_S_k})$ $\times$ TP 
                \State $\{P2P^j_p$ reward$\}_{pub\_S_k}$ = $(\{P^{P2P}_{sup}\}_{pub\_S_k} + \{InDev_j\}_{pub\_S_k} \times \frac{T_{up}}{T_{down}}) \times$ TP $ + \{InDev_j\}_{pub\_S_k} \times (1 - \frac{T_{up}}{T_{down}}) \times$RP
                \State $\{S^{inc}_k\}_{pub\_S_k}$ -= $\{InDev_j\}_{pub\_S_k} \times (1 - \frac{T_{up}}{T_{down}}) \times$ RP
                \EndIf
                \If{$sign(InDev_x) > 0$}
                \State $\{P2P^i_c$ bill$\}_{pub\_S_k}$ = $(\{C^{P2P}_{dem}\}_{pub\_S_k} + \{InDev_i\}_{pub\_S_k} \times \frac{T_{up}}{T_{down}}) \times$ TP $ + \{InDev_i\}_{pub\_S_k} \times (1 - \frac{T_{up}}{T_{down}}) \times$ RP
                \State $\{P2P^j_p$ reward$\}_{pub\_S_k}$ = $(\{P^{P2P}_{sup}\}_{pub\_S_k} + \{InDev_j\}_{pub\_S_k})$ $\times$ TP
                \State $\{S^{inc}_k\}_{pub\_S_k}$ += $\{InDev_i\}_{pub\_S_k} \times (1 - \frac{T_{up}}{T_{down}}) \times$ RP
                \EndIf
                \EndIf
                \If{TD $>$ 0}
                \State $\{S^{inc}_k\}_{pub\_S_k}$ += 0
                \If{$sign(InDev_x) \leq 0$}
                \State $\{P2P^i_c$ bill$\}_{pub\_S_k}$ = $(\{C^{P2P}_{dem}\}_{pub\_S_k} + \{InDev_i\}_{pub\_S_k} \times \frac{T_{down}}{T_{up}}) \times$ TP $ + \{InDev_i\}_{pub\_S_k} \times (1 - \frac{T_{down}}{T_{up}}) \times$ FiT
                \State $\{P2P^j_p$ reward$\}_{pub\_S_k}$ = $(\{P^{P2P}_{sup}\}_{pub\_S_k} + \{InDev_j\}_{pub\_S_k})$ $\times$ TP
                \State $\{S^{exp}_k\}_{pub\_S_k}$ -= $\{InDev_i\}_{pub\_S_k} \times (1 - \frac{T_{down}}{T_{up}}) \times$ FiT
                \EndIf
                \If{$sign(InDev_x) > 0$}
                \State $\{P2P^i_c$ bill$\}_{pub\_S_k}$ = $(\{C^{P2P}_{dem}\}_{pub\_S_k} + \{InDev_i\}_{pub\_S_k})$ $\times$ TP 
                \State $\{P2P^j_p$ reward$\}_{pub\_S_k}$ = $(\{P^{P2P}_{sup}\}_{pub\_S_k} + \{InDev_j\}_{pub\_S_k} \times \frac{T_{down}}{T_{up}}) \times$ TP $ + \{InDev_j\}_{pub\_S_k} \times (1 - \frac{T_{down}}{T_{up}}) \times$FiT
                \State $\{S^{exp}_k\}_{pub\_S_k}$ += $\{InDev_j\}_{pub\_S_k} \times (1 - \frac{T_{down}}{T_{up}}) \times$ FiT
                \EndIf
                \EndIf
                \State $\{S^{bal}_k\}_{pub\_S_k}$ += $\{S^{inc}_k\}_{pub\_S_k} - \{S^{exp}_k\}_{pub\_S_k}$
                \EndFor
                \EndIf
                \If{bid not accepted}
                \For{each $i$, $j$ and $k$ in $C_n$, $P_n$ and $S_n$}
                \State \textbf{goto} Algorithm~\ref{algo:retail_market}
                \EndFor
                \EndIf
                \EndFor
                \EndProcedure
                \end{algorithmic}
        \end{algorithm}
    \begin{itemize}
        \item The total deviation (TD) is calculated as the sum of all prosumer deviations minus the sum of all consumer deviations, equivalent to the difference between the total supply deviation (TSD) and the total demand deviation (TDD).
        \item If the TD is zero, then all users participating in the P2P market buy/sell their actual energy readings at TP, disregarding the specific committed volumes or the individual deviations.
        \item A positive TD indicates that the total volume under-consumed or over-supplied $T\_up = T^c_{under} + T^p_{over}$ is greater than the total volume over-consumed or under-supplied $T\_down = T^c_{over} + T^p_{under}$, total supply being greater than the total demand for electricity. In this case, all downtrenders (over-consumers/under-suppliers) trade their exact energy meter readings entirely at TP, while the uptrenders (under-consumers/over-suppliers) partly compensate for their individual deviations. More specifically, under-consumers buy their committed volumes at TP, sell a proportion of their individual deviation to downtrenders also at TP, and the rest at FiT to suppliers, instead of selling the entire individual deviation at FiT. The over-suppliers sell their committed volumes at TP as usual, but also a share of their respective deviation at TP, before trading the rest at FiT to the RM. These proportions of the sold at TP are based on each respective uptrender's contribution to the total volume under-consumed/over-supplied $T\_up$, and sum up to the total volume over-consumed/under-supplied $T\_down$. 
        \item A negative TD suggests that the energy demand exceeds the energy supply, with the total volume over-consumed or under-supplied $T\_down$ being greater than the total volume under-consumed or over-supplied $T\_up$. In this case, all uptrenders (under-consumers/over-suppliers) buy/sell their final meter readings at TP, regardless of their original committed volumes or deviations, while the downtrenders partially reduce the cost of their individual deviations. In particular, over-consumers purchase their committed volumes at TP as usual, but also a portion of their individual deviation at TP, before buying the rest at RP from the supplier, instead of trading for their entire deficit at RP. Likewise, under-suppliers sell their committed volumes at TP as always, then buy a share of their respective deviation from the uptrenders also at TP, before purchasing the rest of the difference at RP from the RM, instead of making up for their whole individual deviation from trading with the suppliers.
        \item Energy suppliers only trade electricity with the users whose individual deviations point in the same direction as the TD.
        \item Unaccepted bids/offers are settled according to Algorithm~\ref{algo:retail_market}.
    \end{itemize}
    Out of the four billing models presented, this algorithm is the one with the lowest energy volume traded at the RM, which favours the users as the electricity price at the P2P market is advantageous to the one offered by the suppliers. Specifically, the volume traded with the energy suppliers is equal to the absolute value of the sum of all individual deviations of consumers/prosumers:
        \begin{equation}
                V^{RM}_{Univ} = |\sum_{i=1}^{P2P^c_n}InDev_i + \sum_{j=1}^{P2P^p_n}InDev_j | = |\sum_{x=1}^{P2P_n}InDev_x |
        \end{equation}
    \begin{figure}[h]
              \centering
              \includegraphics[width=1.0\textwidth]{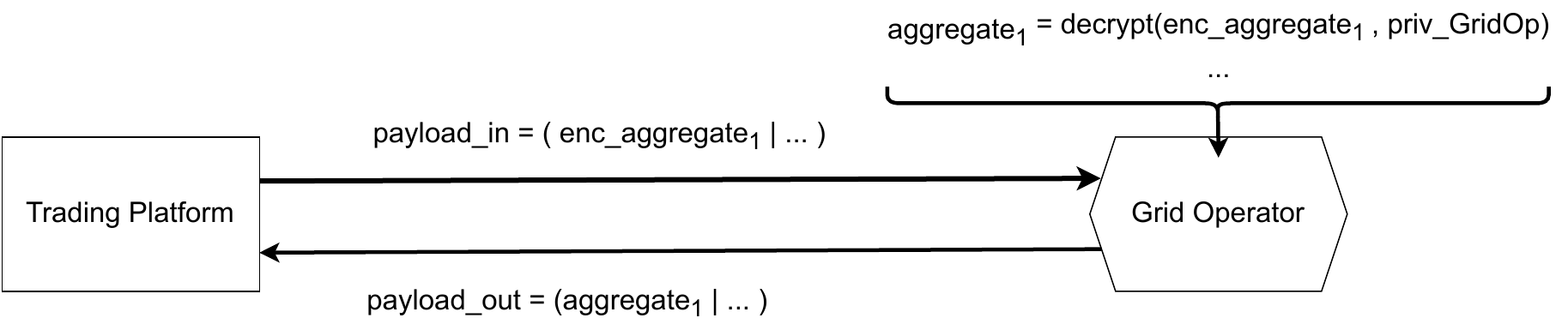}
              \caption{Grid Operator protocol.}
              \label{fig:gridop}
        \end{figure}
    \subsection{Grid Operator Protocol} \label{sec:gridop}
    The grid operator is a trusted third party (TTP) whose responsibility is to decrypt the incoming aggregated values describing the P2P market, to communicate their plaintext form back to the trading platform, and to act as a supervisor of supplier honesty, having the ability to check the final reported bills using the backup values calculated and stored by the trading platform during the billing period. Its protocol is illustrated in Figure~\ref{fig:gridop}. For each trading period, the grid operator's instructions are separated into the following steps, also illustrated in Algorithm~\ref{algo:gridop}:
    \begin{algorithm}[t]
        \caption{Grid Operator Protocol}\label{algo:gridop}
        \begin{algorithmic}[1]
            \setstretch{1.25}
            \Procedure{Decrypted Aggregates}{$payload\_in$}
                \State \hspace{\algorithmicindent}$T^c_{under}$ = $Dec\left(\{T^c_{under}\}_{pub\_GridOp}, priv\_GridOp\right)$
                \State \hspace{\algorithmicindent}$T^c_{over}$ = $Dec\left(\{T^c_{over}\}_{pub\_GridOp}, priv\_GridOp\right)$
                \State \hspace{\algorithmicindent}$T^p_{under}$ = $Dec\left(\{T^p_{under}\}_{pub\_GridOp}, priv\_GridOp\right)$
                \State \hspace{\algorithmicindent}$T^p_{over}$ = $Dec\left(\{T^p_{over}\}_{pub\_GridOp}, priv\_GridOp\right)$
                \State \Return $payload\_out = \left(T^c_{under} \| T^c_{over}\| T^p_{under}\| T^p_{over}\right)$
                \EndProcedure
        \end{algorithmic}
    \end{algorithm}
    \begin{enumerate}
        \item Receive a list of encrypted aggregate values in the form of an incoming payload comprising of the total volume under-consumed, over-consumed, under-supplied, and over-supplied by the P2P users:
            \begin{equation}
                payload\_in = \left(\{T^c_{under}\}_{pub\_GridOp} \| \{T^c_{over}\}_{pub\_GridOp} \| \{T^p_{under}\}_{pub\_GridOp} \| \{T^p_{over}\}_{pub\_GridOp}\right)
            \end{equation}
        \item Decrypt each of the homomorphically encrypted values using the corresponding private key of the grid operator $priv\_GridOp$:
            \begin{equation}
                 decrypt\left(\{X\}_{pub\_GridOp}, priv\_GridOp\right) = X, \text{where } \{X\}_{pub\_GridOp} \in payload\_in
            \end{equation}
        \item Send the plaintext versions of the received values back to the trading platform;
        \begin{equation}
                payload\_out = \left(T^c_{under} \| T^c_{over}\| T^p_{under}\| T^p_{over}\right)
        \end{equation}
    \end{enumerate}
    In addition to these four incoming aggregates, the trading platform also uses the total deviation (TD), total demand deviation (TDD), and total supply deviation (TSD) in some of the billing models (see Sections~\ref{algo:social_split} and~\ref{algo:univ_split}). However, all of them can be derived from the more specific deviation aggregates and should not be redundantly communicated over the network:
    \begin{equation}
                 \text{TDD} = T^c_{over} - T^c_{under}
    \end{equation}
    \begin{equation}
                 \text{TSD} = T^p_{over} - T^p_{under}
    \end{equation}
    \begin{equation}
                 \text{TD} = \text{TSD} - \text{TDD}
    \end{equation}\\
    
    \begin{figure}[h]
              \centering
              \includegraphics[width=1.0\textwidth]{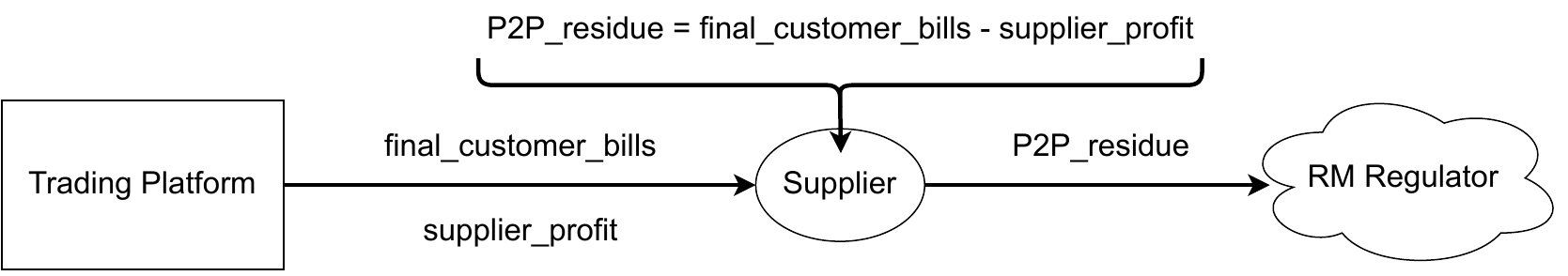}
              \caption{Supplier protocol.}
              \label{fig:supplier}
    \end{figure}
    \subsection{Supplier Protocol} \label{sec:supplier}
        Energy suppliers are responsible for settling the final monthly bills with their contracted customers, as well as ensuring that all payments to/from the P2P market sent/received by these customers, which are included in their final bill, are distributed accordingly across the other suppliers, as illustrated in Figure~\ref{fig:supplier}. The supplier operates in two phases:
    
        \begin{itemize}
            \item Every trading period, the supplier receives its aggregate profit from all customer transactions in that slot $\{S^{bal\_i}_{k}\}_{pub\_S_k}$, which is decrypted using their homomorphic private key $priv\_S_k$:
            \begin{equation}
             decrypt\left(\{S^{bal\_i}_{k}\}_{pub\_S_k}, priv\_S_k\right) = S^{bal\_i}_{k}
            \end{equation}
            From this value, whether negative or positive, they can infer if the market is trending more toward supply or demand, adjusting their offer and preparing for the next trading period appropriately.  
            \item At the end of every billing period, the supplier must resolve the final bills, a process which is separated into multiple steps and illustrated in Algorithm~\ref{algo:supplier}:
            \begin{enumerate}
                \item Calculate total monthly profit using the partial profits communicated throughout the billing period:
                \begin{equation}
                    S^{bal}_{k} = \sum_{i=1}^{no\_slots}S^{bal\_i}_{k}
                \end{equation}
                \item Receive from the trading platform the final electricity bill of each of their customers, which includes both the value these users need to pay to the P2P market and to the RM. They decrypt each of the bills using their homomorphic private key $priv\_S_k$:
                \begin{equation}
                 decrypt\left(\{P2P^x \text{ bill/reward}\}_{pub\_S_k}, priv\_S_k\right) = P2P^x \text{ bill/reward}
                \end{equation}
                \item Carry out the billing process with customers by receiving compensation for the bills from their overall net buyers and paying out the rewards of their overall net sellers of electricity.
                \item Calculate the left-over capital which was traded at the P2P market by subtracting the total supplier profit from the aggregate customer bills:
                \begin{equation}
                    S^{P2P}_{k} = \left(\sum_{i=1}^{P2P^c_{n,k}}P2P^i_c \text{ bill} - \sum_{j=1}^{P2P^p_{n,k}}P2P^j_p \text{ reward}\right) - S^{bal}_{k}
                \end{equation}
                \item Submit the value of the P2P trade residue $S^{P2P}_{k}$ to the retail market regulator in order to facilitate the fair redistribution of the remaining capital among the other suppliers.
            \end{enumerate}
        \end{itemize}
        
        \begin{algorithm}[t]
            \caption{Supplier Protocol}\label{algo:supplier}
            \begin{algorithmic}[1]
                \setstretch{1.25}
                \Procedure{Supplier profit}{}
                   \For{each $i$ in $no\_slots$}
                        \State $S^{bal}_{k}$ += $S^{bal\_i}_{k}$
                    \EndFor
                    \For{each $i$, $j$ in $P2P^c_{n,k}$, $P2P^p_{n,k}$}
                        \State $P2P_c$ bills += $Dec\left(\{P2P^i_{c}~\text{bill}\}_{pub\_S_k}, priv\_S_k\right)$
                        \State $P2P_p$ rewards += $Dec\left(\{P2P^j_{p}~\text{reward}\}_{pub\_S_k}, priv\_S_k\right)$
                    \EndFor
                    \State \Return $ S^{P2P}_{k}$ = $\left(P2P_c\text{ bills} - P2P_p\text{ rewards}\right) - S^{bal}_{k}$
                \EndProcedure
            \end{algorithmic}
        \end{algorithm}
        The intuition behind the left-over balance from the P2P market associated with each supplier is best explained by studying a simple example. Let us assume a system with only one consumer $C_1$ contracted to supplier $S_1$ and only one prosumer $P_1$ contracted to another supplier $S_2$, where $C_1$ buys electricity from $P_1$ at the P2P market. In this case, $S_1$ will receive money from $C_1$ for that volume of energy, while $S_2$ will have to send money to $P_1$ accordingly. What is left is for $S_1$ to pay the appropriate value to $S_2$, and the bill settlement is complete. In a sense, suppliers act as intermediaries for P2P trades. Therefore, without a mechanism for redistributing the capital between suppliers, some would be left with unfair deficits, and others with a surplus of capital.
        \\ \ \\
        After all these left-over balances are submitted to the market regulator, it is checked whether their sum is equal to zero (since the electricity sold must be equal to the electricity bought), indicating that the entire P2P trade has been accounted for and that the values communicated by the suppliers are accurate. In the case of a non-zero value, the grid operator is called upon to verify the calculated values using the backup information stored by the trading platform, encrypted using the grid operator's homomorphic public key. Subsequently, any dishonest supplier is punished accordingly to prevent such behaviour in the future.\\ \ \\
        If the data-gathering step proceeds smoothly, with no inconsistencies detected, the retail market regulator is tasked with managing the fair transfer of funds between energy suppliers, such that all P2P trade residues are sorted out. An example of such a regulator in the UK is Elexon~\cite{elexon-2022}, whose main role is to facilitate the implementation of the Balancing and Settlement Code (BSC), coordinating the settlement processes in the wholesale energy market~\cite{elexon-2020}.

\clearpage
\section{Evaluation} \label{sec:evaluation}
This section assesses the privacy and security properties of the proposed protocol and evaluates it in terms of the computational complexity imposed on different entities in the system, as well as the communication overheads incurred between the various agents.
\begin{figure}[h]
      \centering
      \includegraphics[width=0.8\textwidth]{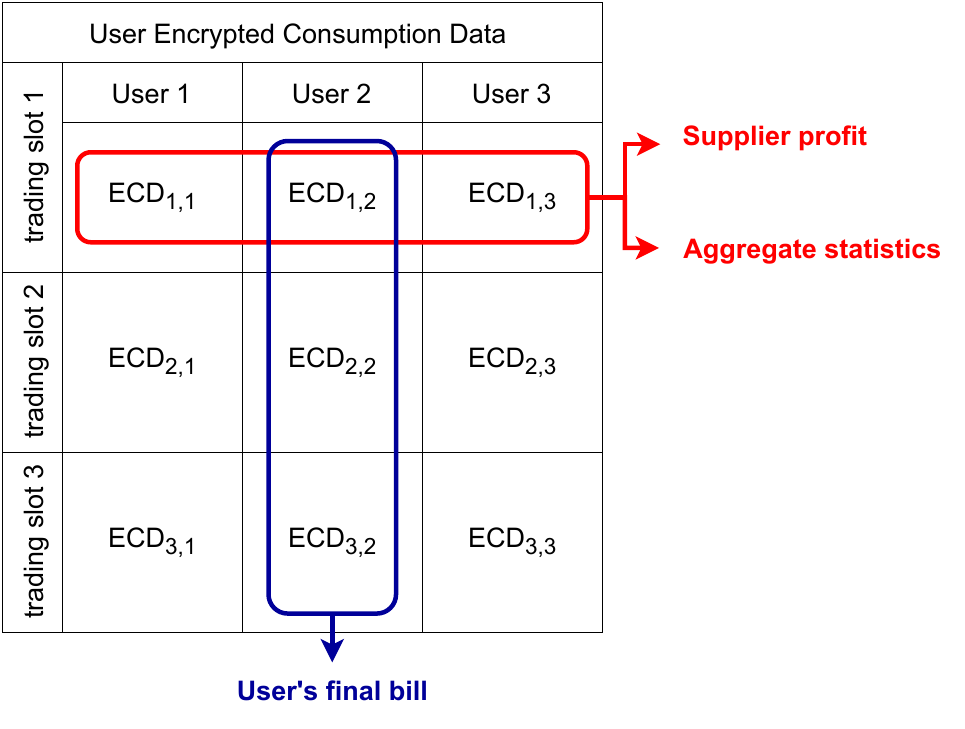}
      \caption{Privacy-preserving use of encrypted user data through aggregation.}
      \label{fig:evaluation}
\end{figure}
    \subsection{Privacy and Security Analysis}
        \subsubsection{Metering Data Confidentiality} 
        Each user's consumption data (CD) is encrypted at its source (SM) using Paillier's partially homomorphic cryptosystem, once using the grid operator's public key and once separately using their contracted supplier's public key. This encrypted consumption data (ECD) is then sent to the trading platform where it is temporarily stored and used in bill calculation in its encrypted format, as the platform does not have access to either of the private keys necessary to decrypt the metering data. A part of each user's ECD, the individual deviation, is aggregated with that of all the other households participating in the P2P market, before being sent to the grid operator for decryption. Therefore, the most fine-grained CDs which the trading platform and the grid operator have access to are the aggregate CDs (ACD) of the entire set of users participating in the local energy market, with the size of this set being on the order of thousands. Thus, even authorised entities such as the trading platform, the grid operator and suppliers do not have access to individual metering data. Figure~\ref{fig:evaluation} demonstrates the aggregation of ECDs for calculating market statistics. In order to also mitigate the risk of data breaches from the trading platform's databases, the individual ECDs are removed from storage after the partial bill calculation for each trading period has ended. Due to the assumed public-key encryption and digital signature infrastructure underlying the communication between authorised entities, the ECD and ACD are also resistant to any eavesdropping attacks whilst in transit between SM-to-TrPlat, TrPlat-to-GridOp, and GridOp-to-TrPlat.
        \subsubsection{Partial Bill Confidentiality}
        The ECD of each user is used in the calculation of the partial bill by the trading platform. Therefore, there are two separate ciphertexts describing the user's bill for each trading period, one encrypted with the homomorphic public key of the grid operator and one encrypted using the public key of the corresponding supplier, with the trading platform being incapable of decrypting either of them as it does not have access to the respective private keys.\\ \ \\
        After every trading slot, the trading platform calculates an aggregate encrypted profit for each supplier using a part of the partial bills of their respective users, which is then communicated to the specific supplier. Because PPBSP follows the ``principle of least privilege", only allowing an entity access to the data it needs to carry out its duties and nothing more~\cite{saltzer1975protection}, the suppliers are not informed of the individual fine-grained electricity bills of their customers. Instead, an energy supplier only receives the aggregate encrypted profit incurred from trading with its customer base (see Figure~\ref{fig:evaluation}), which it can decrypt using its own private key. Assuming a large enough number of users contracted to the same supplier, it is impossible for that supplier to extract any detailed information about a specific user's fine-grained individual deviation, let alone their entire CD.  \\ \ \\
        At the end of each billing period, the partial bills of a user (encrypted with the supplier's homomorphic public key) are summed up into a single encrypted value, which is sent to their respective electricity supplier for bill settlement purposes. Figure~\ref{fig:evaluation} also illustrates this method of aggregation. Assuming half-hourly trading slots over the period of an entire month, this final energy bill would represent the aggregate of at least 1344 partial bills. Therefore, the supplier is unable to deduce any detailed sensitive data from the final communicated bill. Moreover, since the Paillier cryptosystem is semantically secure against chosen-plaintext attacks, as the decisional composite residuosity assumption (DCRA) is considered intractable~\cite{paillier1999public}, and assuming secure and authentic communication channels connecting the system entities, only the corresponding supplier and the grid operator have access to the final monthly bill of any specific user.
        \subsubsection{Supplier Accountability}
        The entire partial bill calculation process performed by the trading platform every trading period is carried out once using the data encrypted by the suppliers' homomorphic public keys, and repeated again using the CDs encrypted with the grid operator's public key. Therefore, instead of each individual supplier being the only one capable of decrypting their users' final bills and being trusted to communicate accurate values to the market regulator, the grid operator also has the capacity to verify any of their calculations using the backup encrypted partial bills. Such, the risk of a supplier deviating from their preimposed protocol is minimised, because of the grid operator's implicit ability to audit their transactions, which would lead to pertinent punishments if found to have behaved in a dishonest manner.

    \subsection{Performance Evaluation}
    The following analysis will focus only on the computational load and communication cost inflicted by the privacy-preserving billing and settlements protocol, as described in Section~\ref{sec:design}, which is exclusively concerned with protecting sensitive user data from authorised entities. As a result, the additional overhead of implementing the assumed secure and authentic communication channels (e.g. established using Transport Level Security~\cite{dierks2008transport}) is omitted.
    \begin{table}[h]
            \caption{Computational complexity of PPBSP}
            \label{table:computation}
            \centering
            \begin{tabular}{l  @{\hskip 0.5in}  p{0.6\linewidth}}
            \toprule
            Entity & Operations per trading period \\
            \toprule
            SM & $4 \times \text{HomoEnc}$ \\
            TrPlat & $\left(2 \times N_u \right) \times \text{BillCalc}$ \\
            GridOp & $4 \times \text{HomoDec}$ \\
            Supplier & $1 \times \text{HomoDec}$ \\
             \toprule
            Entity & Operations per billing period \\
            \toprule
            SM & - \\
            TrPlat & - \\
            GridOp & - / $\left(2 \times N_s \right)\times \text{HomoDec}~^*$ \\
            Supplier & $N_{u,s} \times \text{HomoDec}$ \\
            \bottomrule
            \end{tabular}
            \smallskip
        \parbox[t]{\textwidth}{\footnotesize
        \hskip 0.75in
          $^*$~only on request for inspection.
        }
        \end{table}
        
        \subsubsection{Computational Complexity} 
        Computationally expensive operations used in PPBSP are (homomorphic) key generation, (homomorphic) asymmetric encryption/decryption, and encrypted bill calculation. They are denoted as KeyGen, HomoEnc, HomoDec, and BillCalc respectively. Table~\ref{table:computation} summarises the computational complexity of PPBSP. \\
        \\As described in Section~\ref{sec:design_overview}, PPBSP can be split into three phases:
        \begin{figure}[t]
            \centering
            \subfloat[Linear Time]{\includegraphics[trim={0.05cm 0 0.05cm 0.05cm},clip]{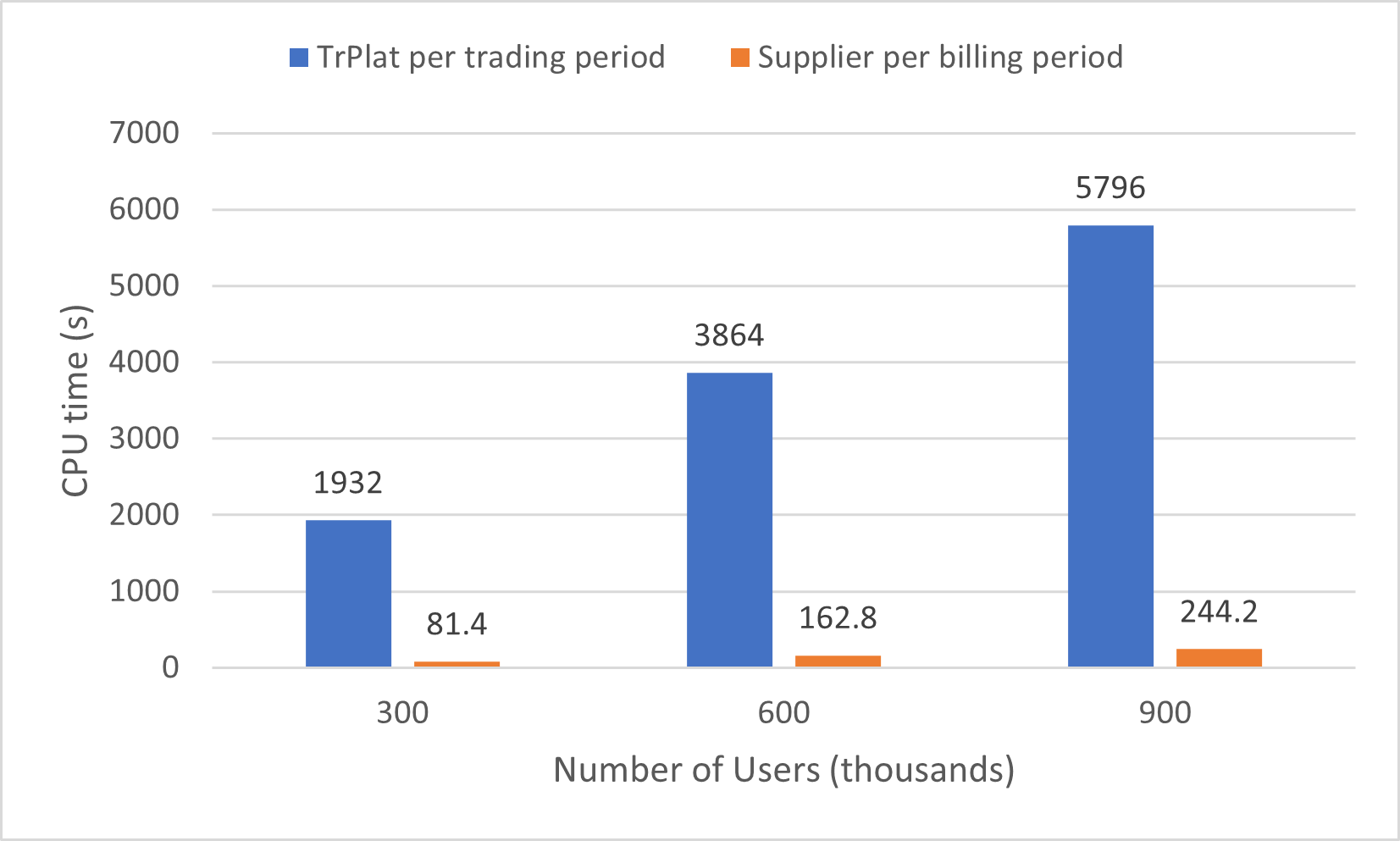}}
            \qquad
            \subfloat[Constant Time]{\includegraphics[trim={0.05cm 0.05cm 0.05cm 0.05cm},clip]{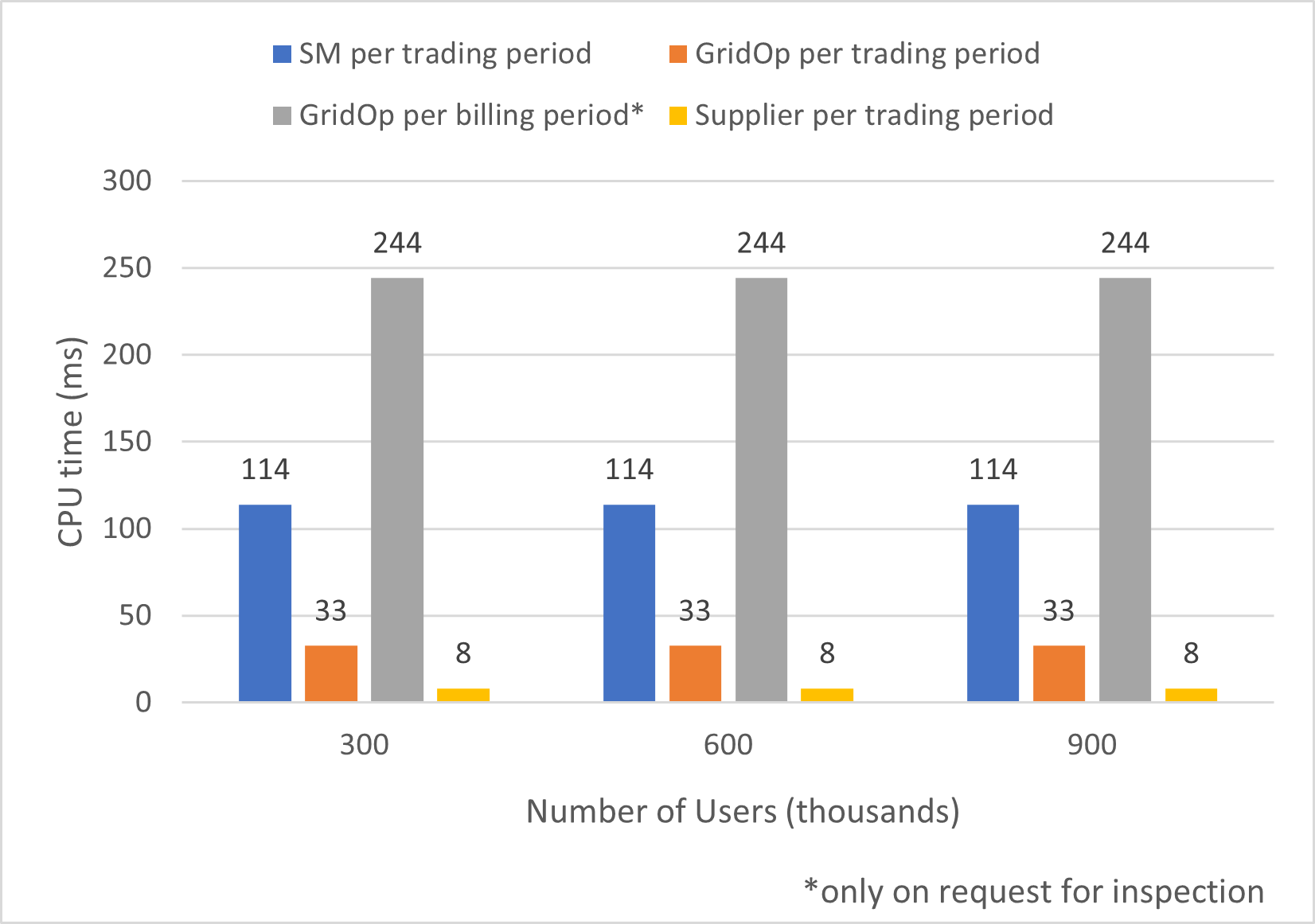}}
            \caption{Computational cost of each PPBSP entity.}
            \label{fig:computation}
        \end{figure}
        \begin{enumerate}
            \item At system initialisation, the grid operator and each supplier generate a pair of public-private keys: 
            $1 \times \text{KeyGen}$
            \item Every trading period (e.g. 30 minutes, one hour, etc.):
            \begin{itemize}
                \item Each SM performs four HomoEnc, encrypting its committed volume and its individual deviation with the respective supplier's homomorphic public key, and separately with the grid operator's public key: 
                $4 \times \text{HomoEnc}$
                \item The trading platform computes two separate partial bills for each user. Crucially, this operation can be parallelised as each partial bill is independent of the others: 
                $\left(2 \times N_u \right) \times \text{BillCalc}$
                \item The grid operator carries out four HomoDec, decrypting the aggregate market statistics received from the trading platform:  
                $4 \times \text{HomoDec}$
                \item Each supplier performs one HomoDec, decrypting their aggregate profit: 
                $1 \times \text{HomoDec}$
            \end{itemize}
            \item Every billing period (e.g. one month):
            \begin{itemize}
                \item The SMs do not perform any computationally expensive operations. 
                \item The trading platform does not carry out computationally intensive calculations either.
                \item The grid operator does not perform any computation unless the suppliers have communicated erroneous results, in which case it aggregates the final customer bills per supplier and then decrypts these aggregates and each of the monthly supplier profits:  
                $\left(2 \times N_s \right)\times \text{HomoDec}$
                \item Each supplier decrypts their customers' bills: 
                $N_{u,s} \times \text{HomoDec}$
            \end{itemize}
        \end{enumerate}
        I also ran simulations on an Intel Core i7-8565U CPU (1.80GHz) with 16GB of RAM in order to demonstrate the scalability of PPBSP. For the experiments, homomorphic encryption was implemented using the python-paillier library~\cite{PythonPaillier} with 2048-bit keys, which meets the current NIST recommendation~\cite{barker2016nist}. The mean run time (over 1000 simulations) of each operation is the following: 
        \begin{itemize}
            \item KeyGen $ =  339.53 \text{ ms} = 0.3395 $ s
            \item HomoEnc $ = 28.48 \text{ ms} = 0.0284 $ s
            \item HomoDec $ = 8.14 \text{ ms} = 0.0081 $ s
            \item BillCalc $ = 3.22 \text{ ms} = 0.0032 $ s
        \end{itemize}
        The computational complexity of PPBSP for a system with 30 suppliers ($N_s = 30$), varying the number of users ($N_u$) from 300k to 900k is plotted in Figure~\ref{fig:computation}. The results illustrate the practicality and scalability of PPBSP in real-world scenarios, especially if the trading platform is hosted on a more powerful machine such as a server cluster, since partial bills are independently computable and thus the entire operation can be parallelised (e.g. 2 CPUs would cut the computation time in half, 4 CPUs to a quarter of the time, etc.). Most importantly, the computational load on the SM, a device that has limited computational resources, is very low.
        \begin{table}[t!]
            \caption{Communication cost of PPBSP}
            \label{table:communication}
            \centering
            \begin{tabular}{l  @{\hskip 0.5in}  p{0.6\linewidth}}
            \toprule
            Protocol segment & Number of bits per trading period \\
            \toprule
            SM-to-TrPlat & $\left(4 \times N_u \right) \times \left(|ciphertext| + |boolean|\right)$ \\
            TrPlat-to-GridOp & $4 \times |ciphertext|$ \\
            GridOp-to-TrPlat & $4 \times |float|$ \\
            TrPlat-to-Sup &  $N_s \times |ciphertext|$ \\
             \toprule
            Protocol segment & Number of bits per billing period \\
            \toprule
            SM-to-TrPlat & - \\
            TrPlat-to-GridOp & - / $\left(2 \times N_s \right)\times |ciphertext|~^*$ \\
            GridOp-to-TrPlat & - \\
            TrPlat-to-Sup & $N_u \times |ciphertext|$ \\
            \bottomrule
            \end{tabular}
            \smallskip
        \parbox[t]{\textwidth}{\footnotesize
        \hskip 0.5in
          $^*$~only on request for inspection.
        }
        \end{table}
        \subsubsection{Communication Overhead}
        The communication overhead introduced by PPBSP can be separated into four distinct parts: data sent from smart meters to the trading platform (denoted SM-to-TrPlat), from the trading platform to the grid operator (TrPlat-to-GridOp), from the grid operator back to the trading platform (GridOp-to-TrPlat), and from the trading platform to suppliers (TrPlat-to-Sup). Table~\ref{table:communication} illustrates the communication cost of PPBSP.
        \begin{figure}[t]
            \centering
            \subfloat[Linear Functions]{\includegraphics[trim={0.05cm 0 0.05cm 0.05cm},clip]{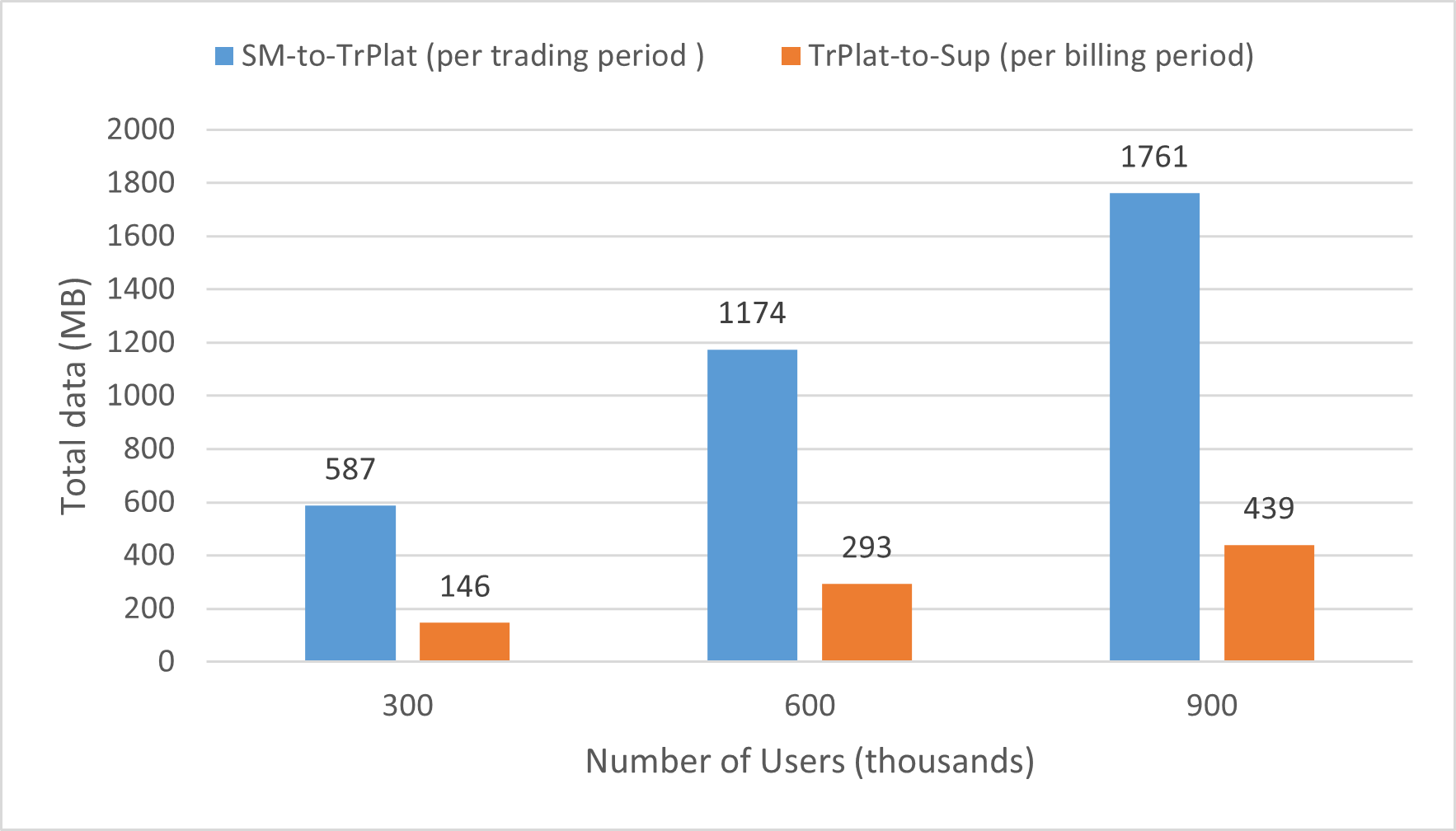}}
            \qquad
            \subfloat[Constant Functions]{\includegraphics[trim={0.05cm 0.05cm 0.05cm 0.05cm},clip]{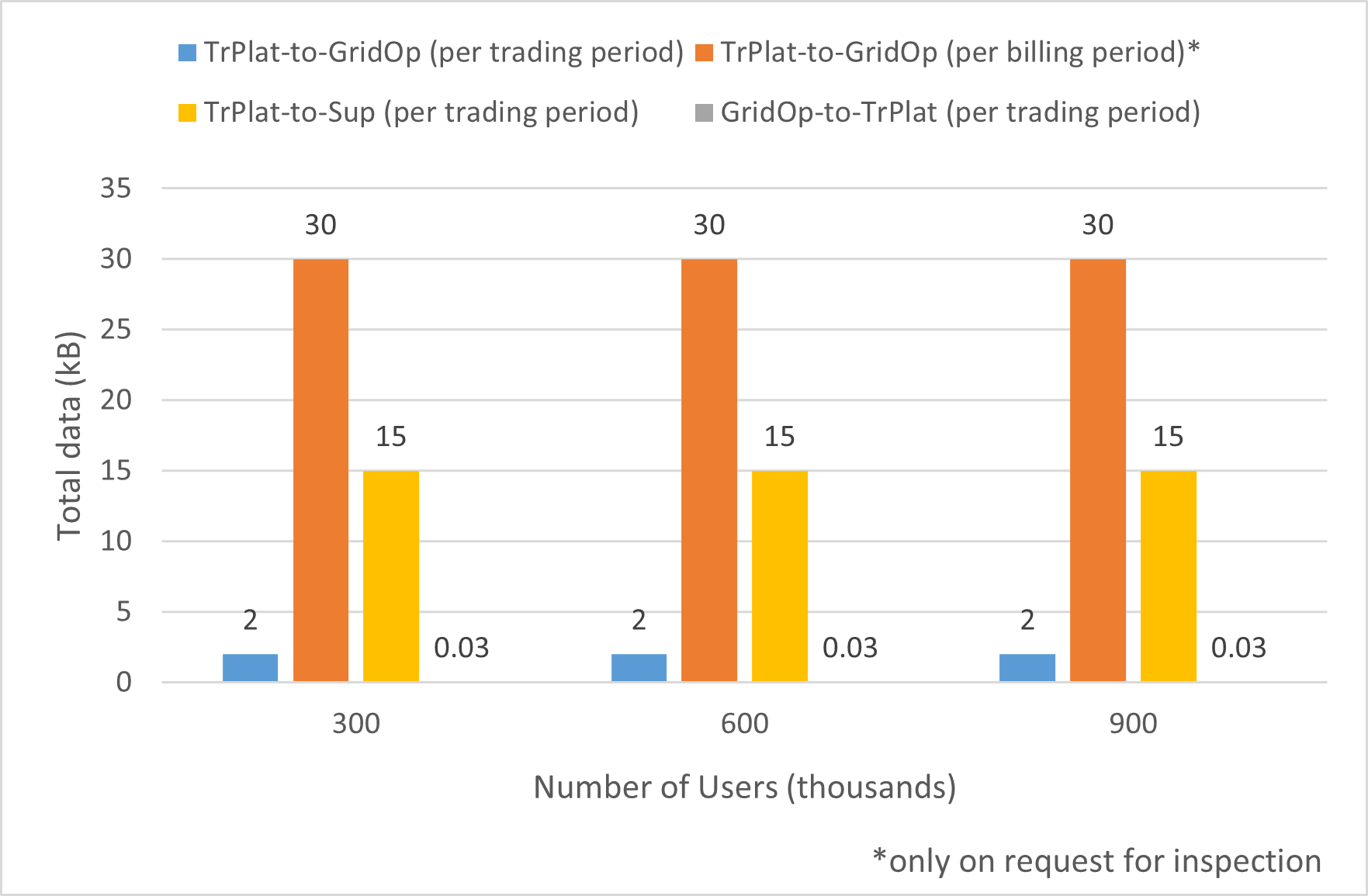}}
            \caption{Communication overhead of PPBSP.}
            \label{fig:communication}
        \end{figure}
        \begin{enumerate}
            \item Every trading period (e.g. 30 minutes, one hour, etc.):
            \begin{itemize}
                \item Each SM sends the trading platform four ciphertexts and four Boolean flags, so the communication cost of SM-to-TrPlat is: 
                $\left(4 \times N_u \right) \times \left(|ciphertext| + |boolean|\right)$
                \item The trading platform communicates four ciphertexts to the grid operator, representing the encrypted market aggregates: 
                $4 \times |ciphertext|$
                \item The grid operator sends back four floating point numbers, which are the decrypted market aggregates:  
                $4 \times |float|$
                \item The trading platform sends each supplier a ciphertext representing their balance change: 
                $N_s \times |ciphertext|$
            \end{itemize}
            \item Every billing period (e.g. one month):
            \begin{itemize}
                \item SMs do not send any information over the network. 
                \item The trading platform does not send the grid operator data unless the suppliers are misbehaving, in which case it sends it the encrypted aggregate bills per supplier and also each supplier's encrypted final profits:
                $\left(2 \times N_s \right)\times |ciphertext|$
                \item The grid operator does not send anything back to the trading platform.  
                \item The suppliers receive from the trading platform their customers' encrypted monthly bills: 
                $N_u \times |ciphertext|$
            \end{itemize}
        \end{enumerate}
    Moreover, I simulate the communication overhead in Figure~\ref{fig:communication}, using the following parameters: $|ciphertext| = 4096 $ bits,  $|boolean| = 8 $ bits (assuming sparse representation of Boolean flags), $|float| = 64 $ bits, the number of suppliers $N_s = 30$, and the user number varying from 300k to 900k.

\clearpage
\section{Conclusions and Future Work} \label{sec:conclusion}
    \subsection{Conclusions}
    In this report, I have designed a novel privacy-preserving billing and settlements protocol for computing and settling bills for users participating in P2P local energy markets. PPBSP uses  partial homomorphic encryption through Paillier's cryptosystem in order to satisfy the LEM's billing requirements in a private manner, while taking into account each user's potential differences between its real meter reading and the electricity volume previously committed at the P2P market auction. The proposed protocol also supports the proportional redistribution of the costs incurred from these deviations among the market's participants, which I have illustrated in the last two proposed billing models. An overview of the entire protocol is presented, before detailing each individual entity's responsibilities and behaviour, as well as explaining the rationale behind these design decisions.\\ \ \\
    Through an informal analysis, I have demonstrated the privacy-protecting properties of PPBSP, fulfilling all of the imposed requirements. Moreover, I have implemented the billing protocol in Python in order to test its performance on a physical, real-world machine. The simulation results and the theoretical cost analysis indicate PPBSP's computational efficiency and the scalability of its communication overheads to realistic-sized P2P markets, especially when considering the contrast in hardware between the powerful trading platform machines and the households' smart meters, devices with very limited computational resources.
	
    \subsection{Future Work}
   In terms of future work, it consists of providing formal proofs of the privacy properties of PPBSP and running large-scale simulations on a real-world SG system, alongside an appropriate trading algorithm, in order to validate the protocol's feasibility. Furthermore, PPBSP could be integrated into the creation of a novel end-to-end protocol for local energy markets, including bid/offer formulation and submission, electricity trading, bill calculation, and bill settlements. \\ \ \\
   The protocol's privacy-preserving characteristics could also be further improved by eliminating the need for communicating homomorphically unencrypted metadata about each household's consumption values to the trading platform, which is currently used in the bill calculation process.

\clearpage
\printbibliography[title={References},heading=bibintoc] 

\end{document}